\documentclass[a4paper,fleqn,usenatbib]{mnras}

\usepackage[T1]{fontenc}
\usepackage{ae,aecompl}

\usepackage{graphicx}	
\usepackage{amsmath}	
\usepackage{amssymb}	

\usepackage{color}

\title[Photochemical haze impact on temperature]{Impact of photochemical hazes and gases on exoplanet atmospheric thermal structure}

\author[P. Lavvas et al.]{
P. Lavvas,$^{1}$\thanks{E-mail: panayotis.lavvas@univ-reims.fr (PL)}
and A. Arfaux,$^{1}$
\\
$^{1}$ Groupe de Spectrom\'etrie Mol\'eculaire et Atmosph\'erique, UMR CNRS 7331, Universit\'e Reims Champagne Ardenne, France \\
}

\date{Accepted XXX. Received YYY; in original form ZZZ}

\pubyear{2020}

\begin{document}
\label{firstpage}
\pagerange{\pageref{firstpage}--\pageref{lastpage}}
\maketitle

\begin{abstract}

We investigate the impact of photochemical hazes and disequilibrium gases on the thermal structure of hot-Jupiters, using a detailed 1-D radiative-convective model. We find that the inclusion of  photochemical hazes results in major heating of the upper and cooling of the lower atmosphere. Sulphur containing species, such as SH, S$_2$ and S$_3$ provide significant opacity in the middle atmosphere and lead to local heating near 1 mbar, while OH, CH, NH, and CN radicals produced by the photochemistry affect the thermal structure near 1 $\mu$bar. Furthermore we show that the modifications on the thermal structure from photochemical gases and hazes can have important ramifications for the interpretation of transit observations. Specifically, our study for the hazy HD 189733 b shows that the hotter upper atmosphere resulting from the inclusion of photochemical haze opacity imposes an expansion of the atmosphere, thus a steeper transit signature in the UV-Visible part of the spectrum. In addition, the temperature changes in the photosphere also affect the secondary eclipse spectrum. For HD 209458 b we find that a small haze opacity could be present in this atmosphere, at pressures below 1 mbar, which could be a result of both photochemical hazes and condensates. Our results  motivate the inclusion of radiative feedback from photochemical hazes in general circulation models for a proper evaluation of atmospheric dynamics.

\end{abstract}




\section{Introduction}

Observations of hot-Jupiters at UV-visible wavelengths reveal a slope in their transit depth. 
While for some of these planets the observed slope is consistent with the anticipated 
molecular scattering, for others the required extinction along the line of sight is much higher, 
suggesting the presence of aerosols ~\citep{Sing16}. Similarly, observations of Neptune-type exoplanets also suggest a variable degree of haziness, with its presence better manifested in transit depth measurements at the 1.4 $\mu$m H$_2$O band \citep{Crossfield17}. However, current data does not allow for a clear distinction between an aerosol layer or a high mean molecular weight atmosphere - a problem that may be ameliorated with future observations with broader wavelength coverage and/or higher precision.
Stellar variability can potentially affect the interpretation of short wavelength observations \citep[e.g.][]{Rackham17}. On the other hand, the persistence of the aerosol-indicator features over multiple star/planet systems suggests that aerosols are an important component of exoplanet atmospheres.

The exact nature of these aerosols is still under debate, with two types of heterogeneous 
components - photochemical hazes and condensates (clouds) - being the two main anticipated families \citep{Morley13, Morley15,Kempton17,Gao20}, 
reflecting our understanding of gas giant atmospheres established through observations of the
solar system \citep{West07,West09}. However, the composition of exoplanet aerosols could be drastically 
different from those in our neighbourhood. In exoplanets condensates include silicates and other exotic 
compounds that are present in the gas phase of these hot environments \citep{Burrows99, Wakeford15, Mbarek16, Woitke18}, while photochemical hazes are expected to have a 
carbon \citep{He18,Fleury19} or sulfur \citep{Zahnle09,Gao17} based structure with possible inclusions of other atoms.  Although the relative 
contributions of condensates and photochemical hazes is not yet fully clarified, the impact of such particles 
on the thermal structure could be significant. 

Previous studies on the impact of aerosols on the atmospheric thermal structure of exoplanets have mainly focused on the role of condensates. Studies on hot-Jupiters considered 
condensates of MgSiO$_3$ and Fe and found that they could have an impact on the thermal structure 
and the emitted thermal radiation, but its magnitude will vary among different planets and according to the 
assumptions made regarding the properties of the cloud particles \citep{Fortney05, Lines18}. Typically, 
the presence of optically thick clouds results in cooling of the atmosphere below the cloud deck 
and heating above, as the altitude range of photon absorption is modified accordingly. 
Nevertheless, the impact of cloud particles on hot-Jupiter's thermal structure seems to be moderate and 
most studies of cloud formation so far have focused on the reverse effect of how horizontal 
temperature variations could affect the cloud formation \citep{Parmentier13,Helling19}. For sub-Neptunes,
studies on GJ 1214b demonstrate that the KCl and ZnS clouds anticipated in this atmosphere \citep{Morley13}
generate a similar effect: temperatures, depending on the cloud particle radius considered, change
by 50 K for 1 $\mu$m particles up to $\sim$300 K for 0.1 $\mu$m particles \citep{Charnay15}. Investigations 
of the microphysical properties of cloud particles suggest radii of a few $\mu$m for silicate clouds in 
hot-Jupiters \citep{Powell18} and KCl/ZnS clouds in sub-Neptunes \citep{Gao18, Ohno18} at the pressure levels where the cloud optical depth is one, implying that a moderate effect is anticipated on the thermal structure from cloud particles in both planet types.  On the other hand, more recent studies find significant radiative feedbacks by clouds on hot-Jupiters \citep{Roman19,Roman20}.

Photochemical hazes are typically produced at lower pressures than those of cloud formation. Microphysics models for such particles suggest particle radii of a few nm in the upper atmosphere, based on the constraints imposed by the observed transit depths and the anticipated formation yields from the 
atmospheric photochemistry \citep{Lavvas17, Gao17, Kawashima18, Kawashima19,Lavvas19, Adams19, Gao20}. 
Depending on their abundance and optical properties photochemical hazes can result in heating or cooling of the atmosphere \citep{McKay91, Arney16, Zhang17}. Previous estimates on the role of photochemical hazes on GJ1214b based on particles with 0.1 $\mu$m radii suggest a significant radiative feedback \citep{Morley15}.
 We explore further this interaction in our study using a radiative-convective model and detailed microphysics modelling of the photochemical haze formation. We focus only on the role of photochemical  hazes, thus do not consider the contribution of cloud opacity in our simulations. 

Most of the previous studies on exoplanet thermal structure have assumed thermochemical equilibrium for the abundances of gaseous species. However, detailed models of disequilibrium chemistry driven by atmospheric mixing and photochemistry \citep{Zahnle09,Moses11,Lavvas14,Venot15} demonstrate that the divergence between the two conditions can be significant for the abundances of the species affecting the thermal structure. An evaluation of the impact of disequilibrium chemistry on the thermal structure suggested temperature differences up to $\sim$100 K \citep{Drummond16}. Thus, we further explore the implications of chemical disequilibrium along with the role of photochemical hazes.

As photochemical hazes are mainly affecting the upper atmosphere's opacity (p < 1 mbar), divergence from local thermodynamic equilibrium (LTE) conditions is anticipated. Here, we do not consider non-LTE effects for the gas phase, which although may be important, are beyond our current goal of evaluation of photochemical haze impact and require a separate dedicated study. However, we do consider the temperature disequilibrium between the particles and the gas envelope that is important in the upper atmosphere \citep{Lavvas17}.
 
Finally, we perform our simulations in 1D so that we can explore in detail the role of photochemical hazes and gases. Nevertheless, dynamics should affect the horizontal distribution of photochemical hazes \citep{Komacek19,Steinrueck20}. Our goal with this study is to evaluate the impact of these components on the thermal structure of the upper atmosphere and motivate the inclusion of their radiative effects in general circulation models of exoplanet atmospheres.

In Section 2 we present a description of the radiative-convective model used in the study (2.1), the opacities  and gas composition considered (2.2), the photochemical haze properties used (2.3) and how the dis-equilibrium between particles and gas temperature is evaluated (2.4). In Section 3 we present results for two reference case exoplanets, those of HD 209458 b (3.1) and HD 189733 b (3.2). As explained further below, for these calculations we consider as input disequilibrium chemical compositions and photochemical haze size distributions that were derived in previous studies, assuming a temperature profile from GCM simulations. In Section 4 we further explore how our simulated temperature profiles impact the interpretation of transit and secondary eclipse observations for these exoplanets, as well as, present a preliminary evaluation of how the revised temperature profiles may affect the disequilibrium chemistry and photochemical haze properties. Section 5 presents our final conclusions.

\section{Methods}

This section provides an overview of the model used in the study, of the gaseous and particulate opacities considered, and how we treat the possible disequilibrium between the particles and the gas background.

\subsection{Radiative-Convective model}

We use a 1D radiative-convective model to simulate the exoplanet thermal structure from the deep atmosphere (10$^3$ bar) to the lower thermosphere ($\sim$10$^{-10}$ bar). RC models provide rapid calculations for the atmospheric thermal structure and have been applied to the Earth's atmosphere \citep{Vardavas84}, atmospheres in the solar system \citep{McKay89, Guillot94, Lavvas08}, as well as to exoplanet atmospheres \citep{Burrows97,Fortney03,Iro05, Amundsen14, Drummond16}. The model used here was described in detail in its previous application to Titan's atmosphere \citep{Lavvas08}, while further details on the radiative transfer methods used can be found in classical textbooks \citep[e.g.][]{Chandrasekhar, Goody, Vardavas07}. As the topic is extensively covered in the existing 
literature, here we provide only a short description of the necessary components.

RC models separate the atmosphere into layers and solve for the vertical temperature profile considering two domains; one where radiative equilibrium dominates the atmospheric energy balance, and one deeper in the atmosphere, where convective mixing provides a more stable energy exchange relative to the radiative equilibrium. In the radiative balance regime, equilibrium requires that the net flux in each atmospheric layer is zero. Thus, the atmospheric temperature is adjusted until the net thermal emission balances the net stellar radiation within each layer. In the convective regime the temperature gradient follows the atmospheric lapse rate, $\Gamma$, and the temperature magnitude adjusts to the value that provides the same energy flux in the transition region, as the pure radiative equilibrium solution would. The location of the transition from a radiative to a convective regime depends on the planet/star case studied and is not certain to always occur within the pressure region under investigation here. 
Thus, we evaluate its presence based on the calculated temperature profile and adjust for it if the simulated temperature profile becomes steeper than the lapse rate. Previous studies suggested that detached convective layers may exist in the high pressure regimes of exoplanet gas-giants \citep{Guillot94, Burrows97}.  However, such transitions occur at higher pressures than those considered here or when the effective temperature of the atmosphere is much smaller than those addressed in our study. Hence, we consider a single convective region in our work.

As implied in the above description, the typical approach for the evaluation of stellar and planetary (thermal) fluxes is to perform independent calculations that provide the flux at each atmospheric level for these two components. For the stellar energy deposition we consider wavelengths ranging between 0.5~\AA~ to 10 $\mu$m, in an expanding grid of 660 wavelength bands. Similarly for the thermal fluxes we consider a grid of 600 wavenumber bands of 50 cm$^{-1}$ width extending from 0 to 30000 cm$^{-1}$. Although the radiative transfer equation can be solved analytically in the case of no scattering \citep{Chandrasekhar}, the contribution of molecular scattering at short wavelengths and haze scattering over the whole spectrum necessitates numerical solvers. Here we use the delta-Eddington approximation and solve for the three moments of the radiative transfer equation for both stellar and thermal radiation, thus allowing 
for the calculation of the net flux at each atmospheric level \citep{Vardavas07}. 

We perform our nominal calculations considering a disk average geometry, thus we divide the incoming stellar radiation by a factor of 4 to take into account the diurnal and horizontal averaging. Although the planets under consideration are expected to be tidally locked to their parent star, we find that disk average insolation conditions are in better agreement with 3D circulation results regarding the thermal structure, compared to cases where the incoming radiation is averaged only over the day-hemisphere (division by factor of 2). For the stellar fluxes simulation we furthermore consider that any radiation reaching the lower boundary is locally absorbed. Practically near the lower boundary both direct and diffuse stellar components attain very small flux values due to the local high opacity. For the thermal radiation fluxes we consider no incoming flux at the top of the atmosphere. At the lower boundary we consider a net outgoing flux characterised by a black body temperature T$_{int}$ that represents energy released from the processes occurring in the interior of the gas giants. Such an addition was found necessary for explaining the observed mass-radius ratios of multiple exoplanets \citep{Guillot02}, with typical values of $T_{int}$ ranging between 100 and 300 K \citep{Iro05,Fortney08}. Subsequent studies suggest that T$_{int}$ may attain higher values for hot-Jupiters \citep{Thorngren19}, however in the simulations presented below we consider $T_{int}$ = 100 K to facilitate comparison with previous thermal structure evaluations.

\subsection{Gaseous opacities}
\begin{figure}
\includegraphics[width=\columnwidth]{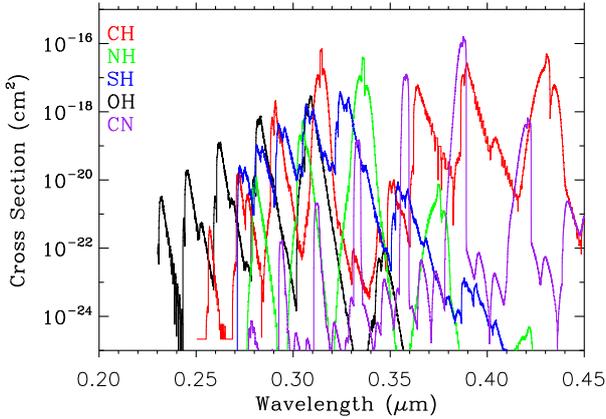}
\caption{Comparison of absorption cross sections of radicals OH, CH, NH, and SH at 1000 K. All cross sections have been smoothed for clarity.}
\label{fig:radical_cs}
\end{figure}

Disequilibrium chemical abundances are calculated with a kinetic model extending from 10$^3$ bar to 10$^{-10}$ bar that solves the continuity equation for each species taking into account atmospheric mixing (eddy), molecular diffusion, and (photo)chemical reactions \citep{Lavvas14}. The eddy profile used is based on general circulation model results \citep{Showman09}.
For the radiation transfer, our gaseous opacity list includes numerous atomic and molecular absorbers the contributions of which varies significantly across the spectrum. At high energies ($\lambda$$<$0.3$\mu$m), our scheme is basically an inheritance from the photochemistry calculations for exoplanet atmospheres, thus it includes absorption cross sections from multiple photochemically active species. At longer wavelengths opacities are calculated based on the correlated-k method \citep{Lacis91} utilising molecular line lists from the Exomol, Theorets, and MoLLIST groups. These include opacities for H$_2$O \citep{Polyansky18}, CO \citep{Li15}, CH$_4$ \citep{Rey17}, NH$_3$ \citep{Coles19}, H$_2$S \citep{Azzam16}, CO$_2$ \cite{Rothman10}, HCN \citep{Barber13}, C$_2$H$_2$ \citep{Chubb20}, TiO \citep[an updated in 2012 version of][]{Plez98}, VO \citep{McKemmish16}, He$_3^+$ \citep{Mizus17} and HeH$^+$ \citep{Amaral19}. In the current simulations we do not treat condensation and cloud formation thus do not include opacities for silicate species.
As our simulations benefit from chemical composition abundances calculated for disequilibrium chemistry, we additionally included opacities for abundant radicals present in the upper atmosphere from the photolysis of major molecules. These include OH \citep{Yousefi18}, CH \citep{Masseron14}, NH \citep{Fernando18}, SH \citep{Gorman19} and CN \citep{Brooke14}. The cross sections of these radicals present strong absorption bands at visible wavelengths (Fig.~\ref{fig:radical_cs}), thus we investigate their possible contribution to the heating of the upper atmosphere. Na and K opacities are calculated as in \cite{Lavvas14}, while bound-free and free-free opacities by H$^-$ and H$_2^-$ are based on \cite{John88}, \cite{Bell87}, and \citep{Bell80}. Furthermore we include Rayleigh scattering by the most abundant species, Thompson scattering by free electrons and collision induced absorption by 
H$_2$-H$_2$ and H$_2$-He pairs. 

\subsection{Haze impact}

Haze particles can efficiently absorb and scatter electromagnetic radiation therefore affect the radiation field in the atmosphere. The degree of this interaction depends on the size distribution of the particles and their optical properties (refractive index). For the particle size distribution we use results from our previous photochemical haze microphysics studies for the atmosphere of HD 189733 b \citep{Lavvas17}, which provided particle size distributions able to satisfy the observed transit depth constraints. In a nutshell, these studies evaluated the particle size distribution assuming a photochemical haze mass flux in the upper atmosphere the magnitude of which was based on photochemistry calculations. The model solves for the particle size distribution (explicitly described in size bins) by solving the continuity equation for each particle size taking into account particle settling, atmospheric mixing (eddy) and particle coagulation and ablation. For the considered photochemical haze optical properties (see below), the particle mass flux is estimated based on the photolysis of the major precursors CH$_4$, HCN and C$_2$H$_2$ assuming a yield for the production of photochemical hazes that is adjusted to fit the observed transit spectra. Further details on the transit spectra simulation and the evaluation of the photochemical haze yield based on observations and laboratory measurements can be found in the relevant studies \citep{Lavvas17, Lavvas19}.

\begin{figure}
\includegraphics[width=\columnwidth]{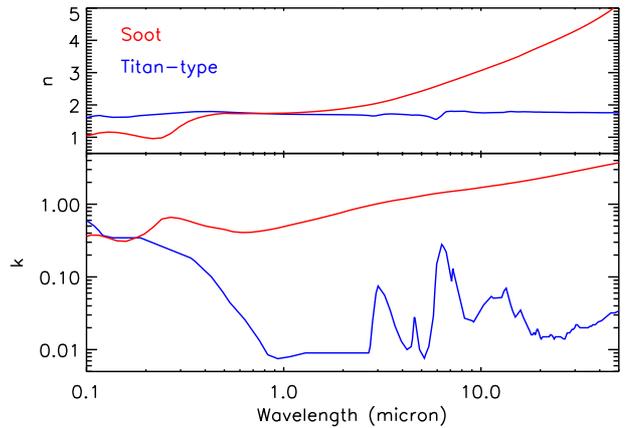}
\caption{Comparison of refractive indices of soot type (red) and Titan-type (blue) compositions. Soots demonstrate a far more absorbing behaviour than the Titan-type composition, manifested over the whole spectrum. On the contrary the Titan-type material absorbs mainly in the short visible and UV wavelengths and at narrow bands in the near IR.}
\label{fig:ref_indx}
\end{figure}
The refractive index of the photochemically produced haze particles in exoplanet atmospheres is currently unknown as the available observational constraints are not sufficient to allow for the retrieval of this parameter, while no laboratory estimations on this parameter have yet been published. Titan's photochemical hazes for which both observational and experimental constraints are available \citep[see review by][]{West13}, are often used as analog for exoplanet photochemical hazes. However, Titan's haze pyrolise at temperatures below 600$^{\circ}$C \citep{Israel05,Morisson16}, which is far lower than the anticipated temperatures in hot-Jupiters. Thus, this type of material cannot 
form or survive at such conditions. A different chemical structure is anticipated for exoplanet photochemical hazes, which would impose further changes in the resulting refractive index of the material. This conclusion leads to the suggestion that a soot-type composition is a more likely representative for hot exoplanets as these materials have a much higher resistance to elevated temperatures \citep{Lavvas17}.  Major differences exist among these two compositions, with the soot composition providing a significantly stronger absorption compared to Titan's analogs, across the whole spectrum (Fig.~\ref{fig:ref_indx}). The properties of exoplanet photochemical hazes may sufficiently change among different planets, reflecting changes in the atmospheric chemical composition and the energy input to the atmosphere \citep{He20}.  As we cannot with certainty constraint this parameter yet, we consider these two (Titan and soot) composition types as limiting cases that potentially envelope the range of photochemical haze compositions in exoplanet atmospheres. 

\subsection{Disequilibrium aspects}
In the interaction of haze with radiation, particles are heated therefore their temperature (T$_p$) increases above that of their surrounding environment (T$_g$). The rate at which the excess particle energy is transferred to the atmosphere depends on the efficiency of collisions of gas species on the particle surface, as well as, the atmospheric opacity at the wavelengths where the excess (thermal) energy of the particles is radiated. The energy balance at steady state for a single spherical particle of radius $r$, takes the form:
\begin{equation}
\pi r^2 \int Q_{\nu}F^{\odot}_{\nu} d\nu - 4\pi r^2 \int Q_{\nu} [\pi B_{\nu}(T_p) - \pi B_{\nu}(T_g)]d\nu - \dot{E}_{col} = 0
\end{equation}
where the first term is the rate at which the particle absorbs stellar radiation, the second is the net rate of energy loss through thermal emission, and the third is the rate of energy exchange through collisions with gas molecules. In the above, Q$_{\nu}$ is the absorption efficiency of the spherical particle, which through Kirchhoff's law, equals its emission efficiency, F$^{\odot}_{\nu}$ is the attenuated stellar flux at the location under investigation, and  B$_{\nu}$ is the black body emission. The collisional term, $\dot{E}_{col}$, depends on the atmospheric density. At locations where the gas mean free path, $\lambda_g$, is much larger than the particle radius (free molecular regime, $Kn$=$\lambda_g$/r$>>1$) and the particle velocity is small compared to the thermal velocity of the gas, the exchange rate is given by \citep{Gombosi}:
\begin{equation}
\dot{E}_{col} = \alpha_T \pi r^2 \left(\frac{\gamma+1}{\gamma-1}\right)\frac{1}{2}p V_T\left(1-\frac{T_p}{T_g}\right)
\end{equation}
where $\alpha_T$ is the thermal accommodation coefficient (assumed unity in our calculations), $\gamma$ the heat capacity ratio, p the gas pressure and V$_T$ the gas thermal velocity. At the opposite limit of continuous flow ($Kn<<1$), the energy exchange rate is described through the thermal conduction, $\kappa$, of the gas:
\begin{equation}
\dot{E}_{col} = 4\pi r \kappa (T_p-T_g)
\end{equation}
There is no analytical solution able to describe the transition between these two regimes. Instead, various approximations have been suggested. For example, \cite{Sherman63} suggested a simple harmonic mean formula $\dot{E}$ =  $\dot{E}_c \dot{E}_f$/($\dot{E}_c$+$\dot{E}_f$) to describe the transition region, while similar approaches where further suggested by other researchers based on experimental measurements \citep[see review by][]{Filippov99}. Here, we use an interpolation formula given by \citep{Pruppacher}:
\begin{equation}
\dot{E} = 4\pi r \frac{\kappa}{\delta} (T_p - T_g) ~\quad with ~\quad \delta=\frac{r}{r + \lambda_g} + \frac{V_T\kappa}{\alpha_T r \rho c_p}
\end{equation}
which converges to the above two limiting cases at the appropriate conditions of $Kn$ values. For small temperature differences between the particle and the gas the interpolation formulae are in agreement with detailed stochastic simulations \citep{Filippov99}. 

\begin{figure}
\includegraphics[width=\columnwidth]{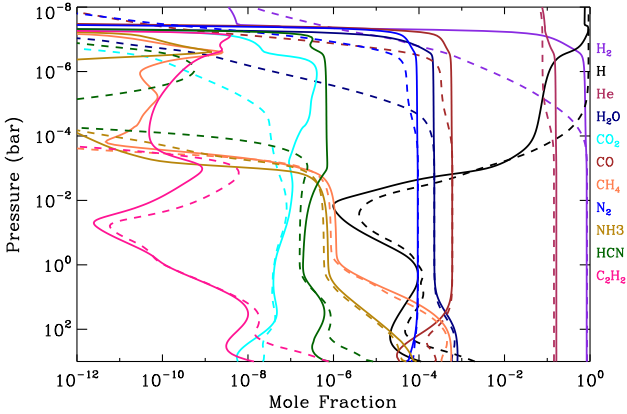}
\caption{Dis-equilibrium chemical composition for the atmosphere of HD 209458 b assuming the \citet{Showman09} day-average GCM temperature profile (dashed lines) and the dis-equilibrium composition temperature profile calculated in this work for complete (4$\pi$) redistribution (solid lines).}
\label{fig:HD209_comp}
\end{figure}

\begin{figure*}
\includegraphics[scale=0.35]{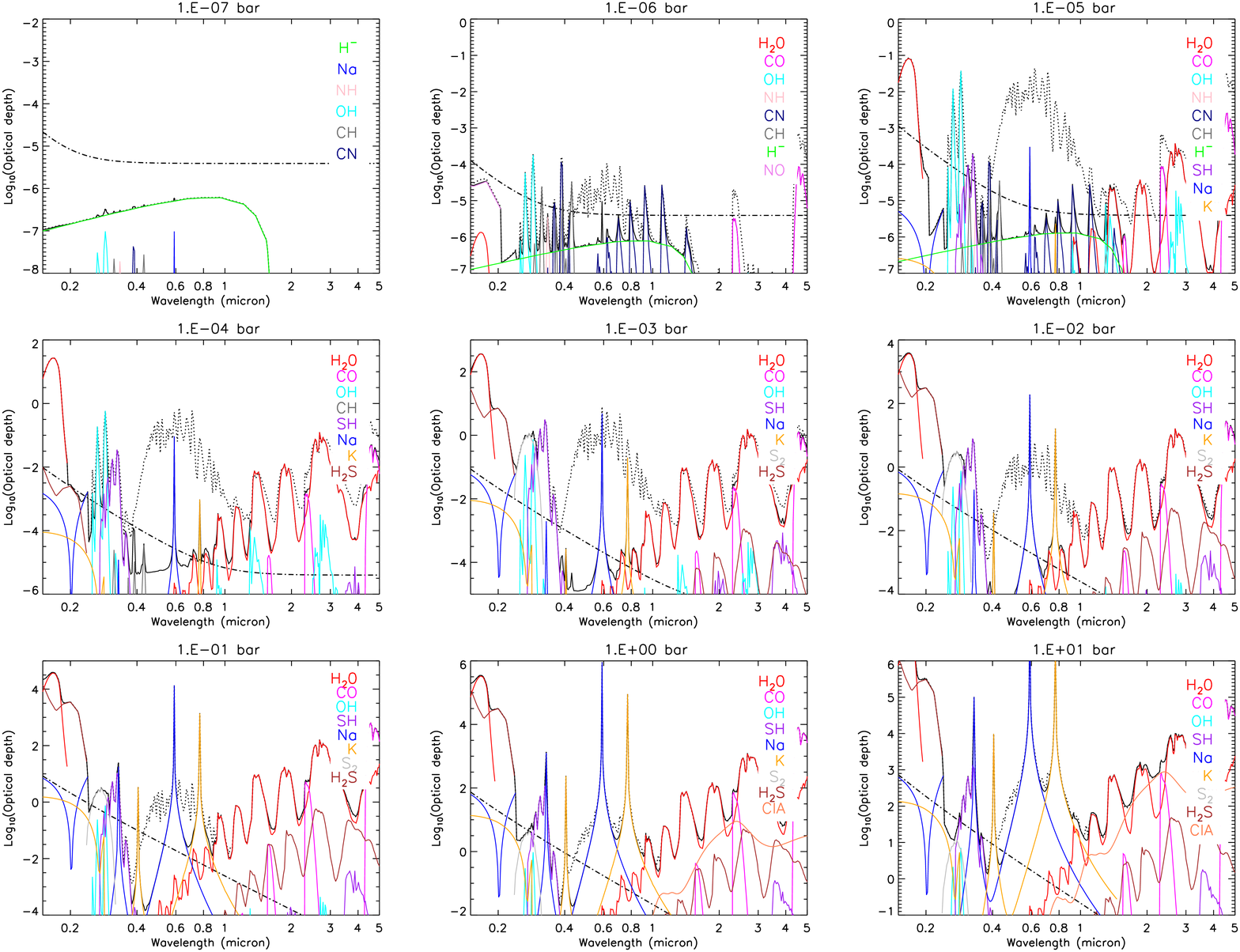}
\caption{Gaseous opacity contributions across wavelength for stellar energy deposition calculations and at different  pressures in the atmosphere of HD 209458 b. Different colours present the absorption optical depth contribution by different gases as labeled in each panel. The solid black line corresponds to the total absorption optical depth based on the dis-equilibrium composition, while the dotted line is the total opacity when the contributions of TiO and VO (evaluated at thermodynamic equilibrium) are also taken into account. The dash-dotted lines present the total scattering opacity (molecular and electrons).}
\label{fig:opacity_gas}
\end{figure*}

The above equations allow for the evaluation of the particle equilibrium temperature at different atmospheric levels. As the gas collision term becomes less efficient with decreasing pressure, particles warm up and their excess energy is balanced by their thermal emission. Previous evaluations for hot-Jupiters have demonstrated how the particle temperature starts to diverge from the gas temperature at pressures below $\sim$10 mbar and with increasing magnitude as particle radius increases. Thus, in the middle and upper atmosphere the particle heating efficiency decreases significantly below unity \citep{Lavvas17}. However, using this particle heating efficiency as a multiplicative factor of the haze extinction could be misleading for the evaluation of the atmospheric temperature. Even if the collisional energy transfer to the gas phase is small, stellar photons at different energies are absorbed by the haze, thus are not available to heat deeper levels of the atmosphere. In addition, the thermally emitted photons from the particles could be absorbed by the gas (and haze) envelope at different atmospheric locations instead of escaping the atmosphere, therefore could further lead to heating. 

To properly address the impact of disequilibrium between the gas and particle temperatures on the thermal structure we included the excess radiative energy of the hot particle in the radiative transfer scheme and simulated the thermal energy fluxes deposited at different atmospheric levels only from this component. The source function at each atmospheric level is calculated from:
\begin{equation}
S_{\nu} = \frac{1}{\chi_{\nu}}\sum_p [B_{\nu}(T_p)-B_{\nu}(T_g)]\sigma_p n_p
\end{equation}
with $\chi_{\nu}$ the local extinction coefficient and $\sigma_p$ and $n_p$ the emission cross section and number density of size bin $p$ particles. The calculated thermal fluxes from the particles are subsequently added to the solar and planetary components for the evaluation of the net flux.

\section{Application to reference cases}

Here we describe the application of the model to the reference exoplanet atmospheres of HD 209458 b and HD 189733 b. As described above we use as input for the gas and photochemical haze  composition results from our previous evaluations. However, as explained below, these studies considered as input thermal structure profiles derived from independent evaluations that assumed chemical equilibrium and clear atmosphere conditions (i.e. no hazes). Therefore, the results presented below serve to demonstrate the implications of disequilibrium chemistry and photochemical haze opacity on the thermal structure and to motivate the need for a self-consistent description in the future. 

\subsection{HD 209458 b}

We start with the case of an apparently clear atmosphere, that of HD 209458 b. This planet has been a paradigm of exoplanetary investigations and has been the focus of multiple studies of thermal structure due to availability of observations \citep{Guillot02,Iro05,Fortney06,Showman09}. Therefore, it provides an opportunity to validate our model against observations and compare with previous studies, before investigating the role of photochemical hazes.

HD 209458 is a G0V star with similar spectral characteristics to the sun. Thus, we consider a similar spectrum to the sun scaled to the stellar radius of HD 209458 for the stellar insolation. For the atmospheric chemical composition, we consider our previous evaluation of disequilibrium composition \citep{Lavvas14} that assumed solar elemental abundances. These computations where performed using the day-average temperature profile evaluated from a general circulation model \citep{Showman09}. For clarity these inputs are presented as dashed lines in Figs.~\ref{fig:HD209_comp} \&~\ref{fig:temp209}.

\begin{figure}
\includegraphics[width=\columnwidth]{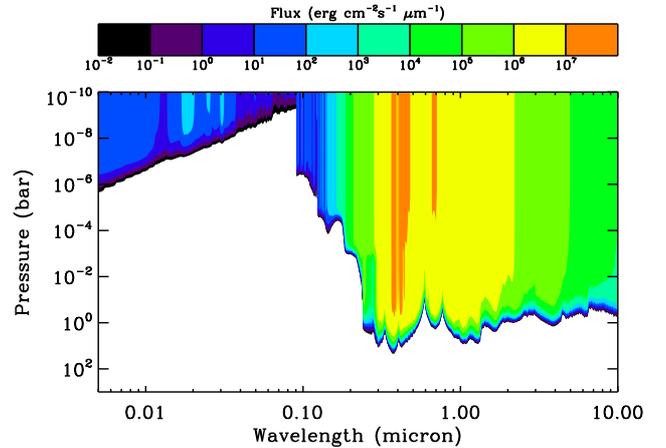}
\caption{Stellar flux penetration in the atmosphere of HD 209458 b at different wavelengths. }
\label{fig:flx_str_209}
\end{figure}

A demonstration of the different species contributing to the atmospheric optical depth at different pressure levels reveals the major opacity sources (Fig.~\ref{fig:opacity_gas}). We focus here mainly on the wavelength range relevant to the deposition of stellar energy. At pressures below 1$\mu$bar and $\lambda$$<$0.1$\mu$m the atmospheric opacity is driven mainly by H$_2$O and atomic/molecular hydrogen (not shown), while at longer wavelengths we find that H$^-$ provides the strongest absorption contribution, which however has a small optical depth ($\tau$$<$10$^{-6}$) and is smaller than the Thompson scattering opacity by electrons. Near 1 $\mu$bar, the densities of primary radicals (dominated by OH, CH and CN) become significant enough to provide a clear signature in the opacity at visible wavelengths (however they remain too small to be observable), while in the near IR the CO contribution is evident. Deeper in the atmosphere, at 10 $\mu$bar, the absorption bands of H$_2$O in the near IR start to have a clear signature in the total opacity, while the Na and K signatures in UV (photoionization) and visible also clearly appear. 

\begin{figure}
A\\
\includegraphics[width=\columnwidth]{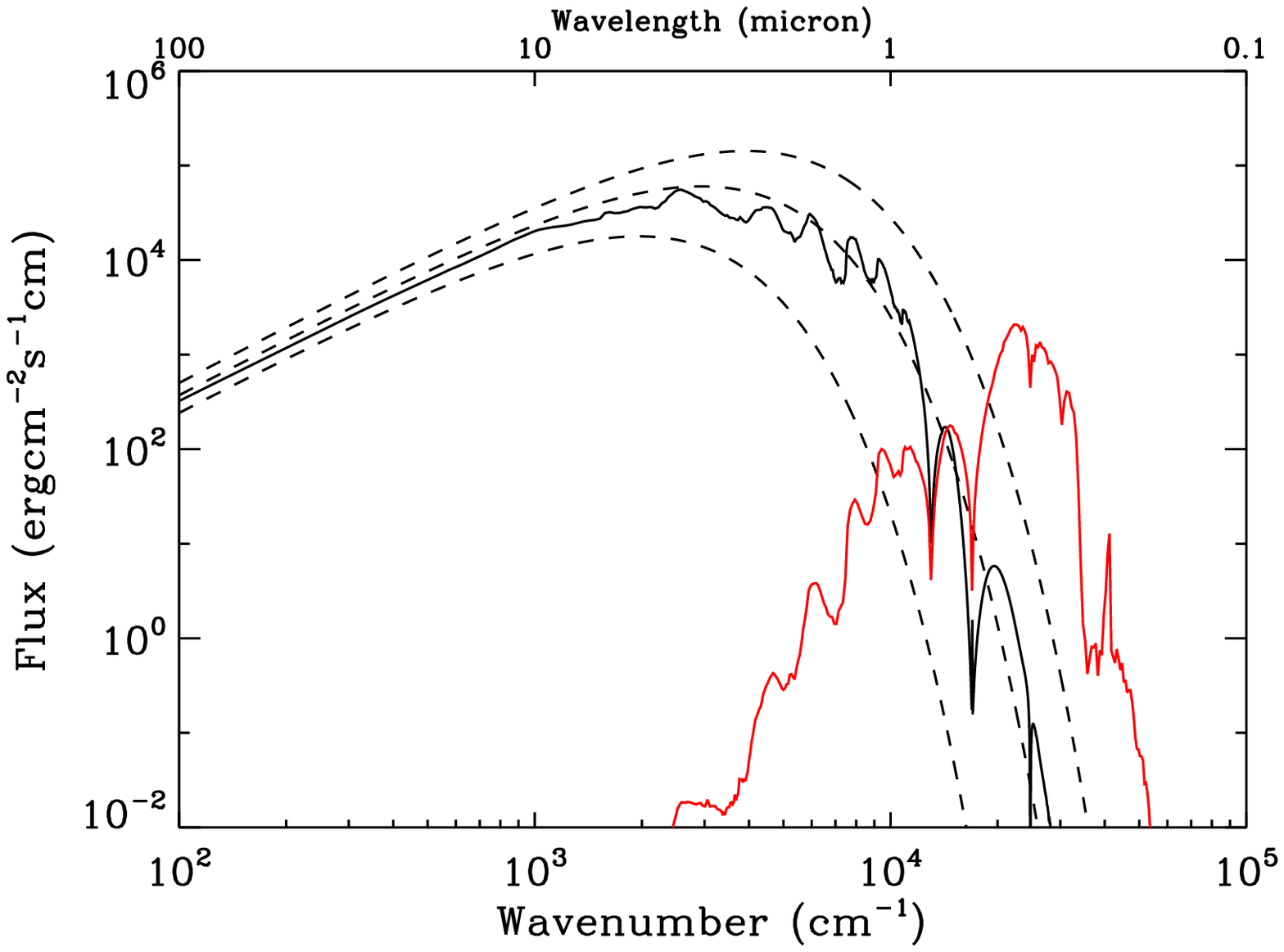}
B\\
\includegraphics[width=\columnwidth]{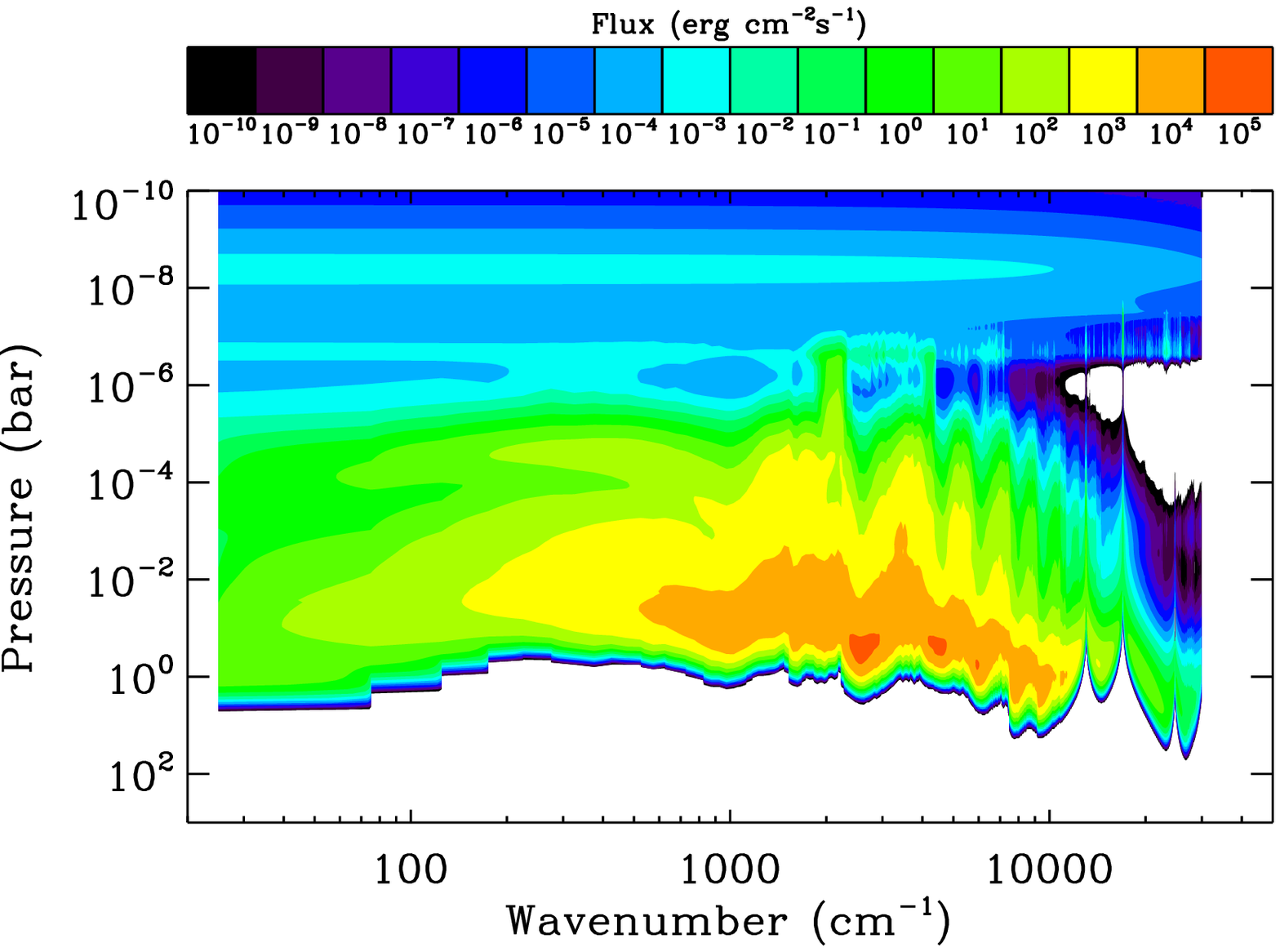}
\caption{Top: Fluxes escaping the atmosphere of HD 209458 b as thermally emitted radiation (black line) and scattered stellar radiation (red). The dashed lines correspond to black body emission at 1000 K, 1500 K and 2000 K. Bottom: Contribution of different pressure levels to the total emitted radiation at the top of the atmosphere across different energies.}
\label{fig:flx_209}
\end{figure}

At higher pressures, another radical that starts to contribute near 0.1 mbar is SH that is formed from the photolysis of H$_2$S. The contribution of SH was considered potentially important according to the preliminary evaluations by \cite{Zahnle09}. However, their SH cross section was based on reported line widths from the solar chromosphere that where subsequently revised and the novel constraints to the SH line profiles results in a significantly weaker cross section \citep[by factors $>$10 at visible wavelengths][]{Gorman19}. Nevertheless, even with the revised cross section SH does have a dominant contribution to the opacity near 0.3 $\mu$m at pressures below $\sim$0.1 mbar.  Similarly, S$_2$ is another major opacity source below 1 mbar, centred in the boundary between UV and visible radiation. Previous studies suggested the theoretical cross section evaluated by \cite{Heijden01}. Yet these calculations correspond to high pressure plasma conditions that are not representative of the relevant atmospheric conditions under investigation here. Instead, we use a cross section evaluated from an experimentally derived linelist and applied to higher temperatures appropriate to exoplanet conditions (R.V. Yelle personal communication). 

Moving to deeper pressure levels of HD 209458 b's atmosphere the main opacity sources remain the same but their individual contributions vary. Near 1 bar H$_2$O dominates over the near-IR, Na and K in the visible and SH, S$_2$ near 0.3 $\mu$. Contributions by H$_2$S are also important, while at p > 1 bar opacity by collision induced absorption starts to be significant. 

\begin{figure*}
\includegraphics[width=\columnwidth]{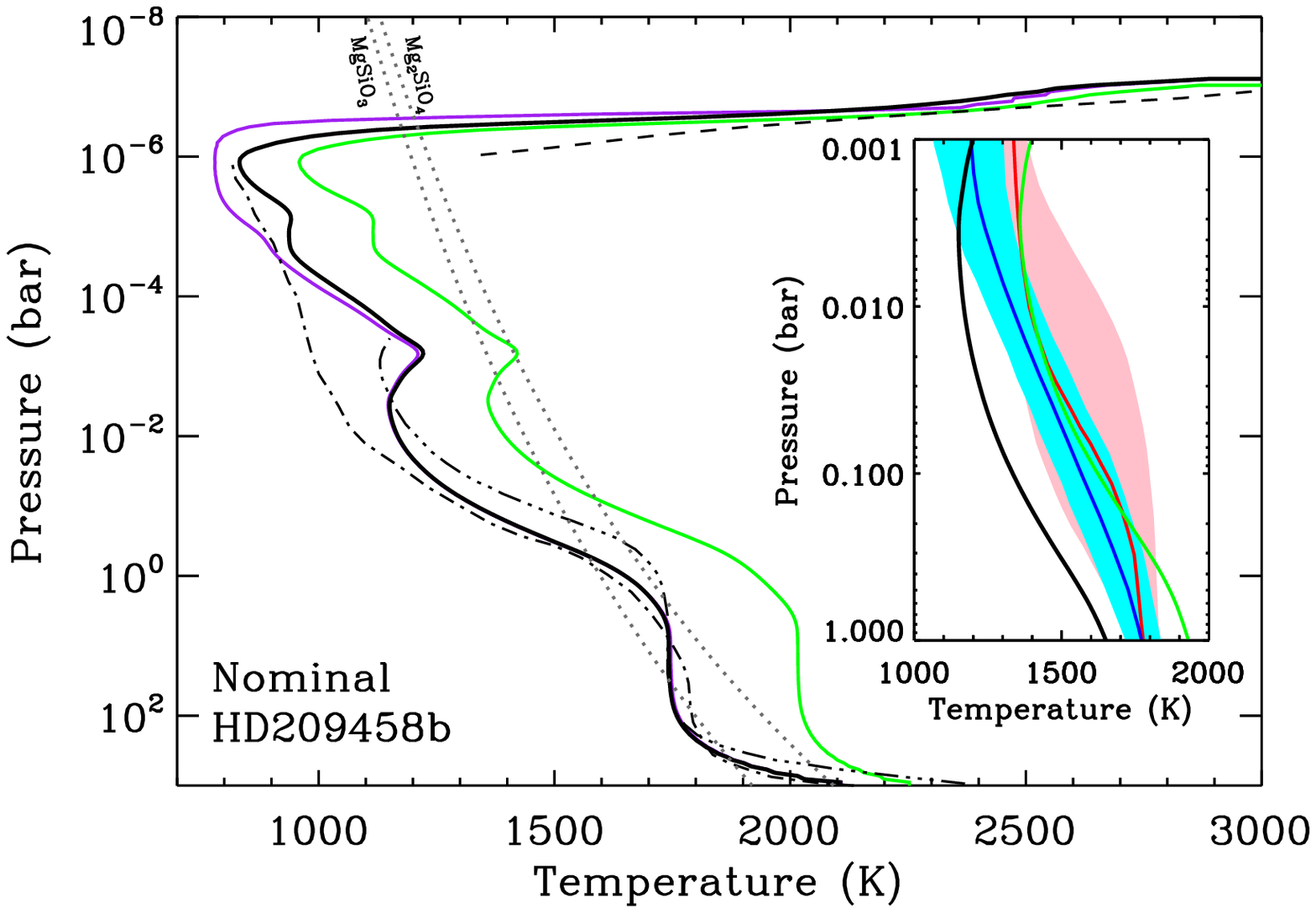}
\includegraphics[width=\columnwidth]{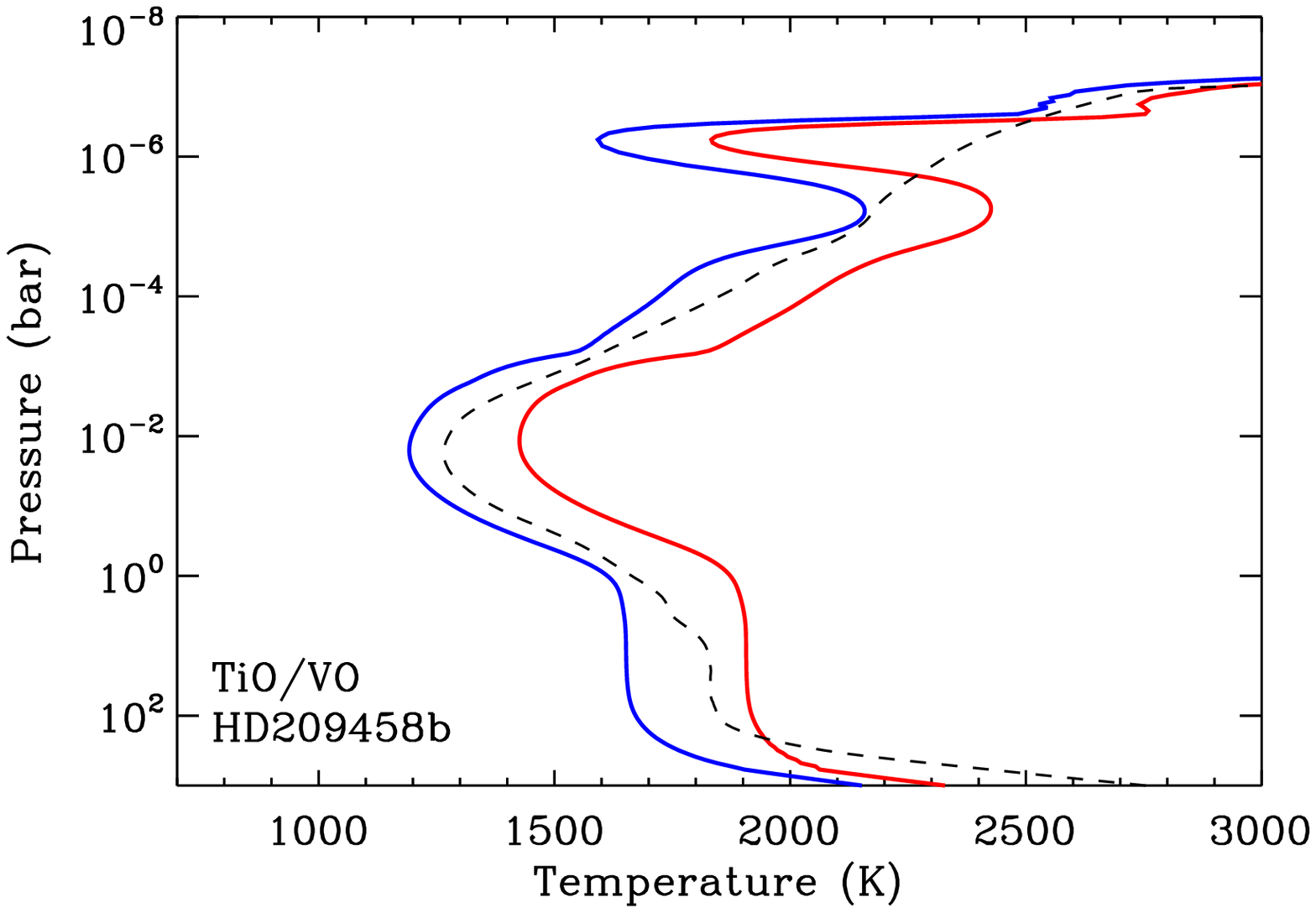}
\caption{Temperature profiles for HD 209458 b. Left: Temperature profile under nominal conditions of disequilibrium chemistry (thick solid black line) compared to other 1D simulations assuming thermochemical equilibrium composition \citep[dash-dotted,][]{Iro05}, \citep[dash-triple-dotted][]{Fortney06}. The purple line presents our simulated temperature profile without the contribution of photochemical radicals in the simulations (see text). All these simulations assume a 4$\pi$ redistribution of incoming stellar energy and T$_{int}$=100 K. The green line presents our nominal thermal structure assuming energy deposition on the day side only (2$\pi$ redistribution). The black dashed line presents the thermospheric temperature profile from escape models \citep{Koskinen13}, and the grey dotted lines the equilibrium saturation curves for MgSiO$_3$ and Mg$_2$SiO$_4$ \citep{Visscher10}. For all simulations, convection does not occur in the simulated pressure range. The inset compares our nominal 2$\pi$ (black) and 4$\pi$ (green) temperature profiles with the profiles retrieved by \citet{Line16} based on secondary eclipse observations. The red and blue profiles with associated uncertainties are based on different parameterizations for the temperature structure used in their analysis. Right: Temperature profiles including TiO/VO at equilibrium density (see text). The thick lines present our simulations for 4$\pi$ (blue) and 2$\pi$ (red) stellar energy redistribution and the dashed line corresponds to the day-average temperature profile from 3D simulations \citep{Showman09}, which is the temperature profile used for the evaluation of the disequilibrium chemistry.}
\label{fig:temp209}
\end{figure*}

The above opacity distribution imposes a direct signature to the stellar fluxes penetrating the atmosphere (Fig.~\ref{fig:flx_str_209}). At $\lambda$ $<$ 900~\AA ~(the H ionization limit) photons are absorbed at very low pressures (p < 1$\mu$bar), while at longer wavelengths the H$_2$O, SH, and S$_2$ define the penetration depths in the FUV. Photons at these energies reach progressively deeper pressure levels down to $\sim$1 mbar at 0.2 $\mu$m and $\sim$1 bar at 0.3 $\mu$m.  At visible wavelengths Na and K dominate the penetration of radiation, while at near-IR H$_2$O dominates the extinction of the stellar fluxes with stellar photons completely absorbed by $\sim$10 bar.

For the planetary emission, the corresponding thermal opacity is dominated by H$_2$O at most IR wavelengths, while as discussed above, Na and K opacities dominate at visible. The planetary emission spectrum is characteristic of a black body of $\sim$1500 K with clear signatures of H$_2$O bands and Na and K absorption at the wings and emission at the cores of the atomic lines (Fig.~\ref{fig:flx_209}). Our results demonstrate that the majority of the outgoing thermal radiation originates from pressure levels between 1 bar and 1 mbar (Fig.~\ref{fig:flx_209}). The situation only changes at the cores of the alkali lines which are optically thick at low pressure levels and their emission is significantly affected by the upper atmosphere temperature. The scattered stellar light flux leaving the atmosphere (Fig.~\ref{fig:flx_209}) becomes comparable to the thermally emitted radiation near the K line and exceeds the thermal component at shorter wavelengths.

The balance between the stellar and planetary fluxes results in the simulated thermal profile (Fig.~\ref{fig:temp209}). We first compare our calculations with previous 1D studies considering thermochemical equilibrium composition with condensation effects included in the abundances of gaseous components. Specifically we compare with profiles generated under the assumption of complete redistribution (4$\pi$) of the incoming radiation and an internal temperature boundary condition of T$_{int}$=100K \citep{Iro05,Fortney06}. Our simulated temperature profile is similar to the previous evaluations in the lower atmosphere demonstrating that differences in the chemical composition between equilibrium and dis-equilibrium conditions do not have a major impact on the thermal structure below $\sim$ 0.1 bar for this planet.  

At lower pressures however, the contribution of SH and S$_2$ visible opacities based on the disequilibrium chemistry induce a significant atmospheric heating between 10 mbar and 0.1 mbar, which was not present in previous evaluations. Without these opacities, the temperature near 1 mbar drops by $\sim$ 100K relative to the nominal profile, thus approaching further the \cite{Iro05} profile. At pressures below 1$\mu$bar our profile demonstrates a sharp temperature increase with temperatures reaching values above 2500 K, which brings our simulated profile close to evaluations of the thermospheric temperature structure \citep{Koskinen13}. This rise reflects the loss of the main atmospheric coolant, H$_2$O, at these pressures. Our simulation does not include all the involved processes at these low pressures (e.g. cooling by advection and radiative recombination) for our results to be directly compared to the escape simulations. We note however that the radicals produced by the photolysis of the main species do affect to a small degree the thermal structure in the lower thermosphere. Our calculations suggest that removing their opacity contribution results in a maximum $\sim$120 K drop in the atmospheric temperature near 10$\mu$bar relative to the nominal profile. Thus the contribution of these species, as well as, of other radiative active molecules may affect the evaluation of the lower thermosphere conditions in escape simulations.

We also compare our calculations with the results from 3D simulations. For this comparison we consider the day-average profile derived from the GCM results of \cite{Showman09}. The latter authors considered a thermochemical equilibrium composition but furthermore included the abundances of TiO and VO without any condensation effects.
The contribution of TiO and VO is particularly important in the upper atmosphere as there the equilibrium abundances demonstrate a maximum, which in combination with the strong absorption bands at visible wavelengths for these molecules facilitates a strong local energy deposition. This effect is demonstrated by the dotted line in Fig.~\ref{fig:opacity_gas}. As anticipated, considering the TiO-VO equilibrium densities based on the GCM temperature profile along with the disequilibrium chemical composition for the other species, our simulated temperature demonstrates a strong increase in the upper atmosphere with a corresponding cooling of the lower atmosphere, relative to the nominal profile (Fig.~\ref{fig:temp209}). The resulting thermal structure is rather similar to the GCM results. Our simulated profile assuming complete redistribution is slightly cooler than the day-average results from the GCM simulation. Assuming that the energy is deposited only on the day side (2$\pi$) we obtain a profile that is substantially warmer than the GCM results. Thus, we conclude that assuming a complete redistribution for the 1D simulations provides more representative results for the average temperature profile. Nevertheless, as we discuss below, the secondary eclipse observations are better reproduced assuming energy deposition on the day side.

Note that the thermal structure below 10 bar would require an extensively long simulation time for convergence, which is unrealistic for a GCM. Thus, the temperature profile there is forced to converge to an 1D simulation \citep{Fortney06}, which however is warmer from the 1D results presented in the nominal case. This suggests that the GCM profile at p > 10 bar probably corresponds to a higher T$_{int}$ value, thus cannot be directly compared to our results. Moreover, the GCM simulations do not extend to pressures smaller than 2$\mu$bar and the presented profile was interpolated with a hydrodynamic escape simulation temperature profile \citep[see][for details]{Moses11}. The local minimum in our simulated profile near 1$\mu$bar is strongly affected by CO emission.

Our results demonstrate that the differences between equilibrium and disequilibrium composition can have major implications for the thermal structure of the upper atmosphere, thus affect the local chemical composition. We explore these implications further in section 4.

\begin{figure}
\includegraphics[width=\columnwidth]{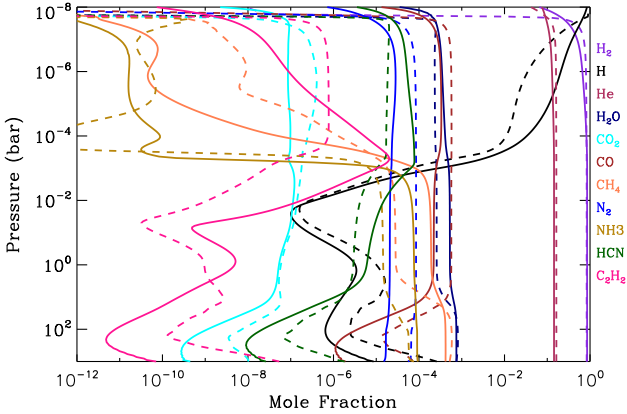}
\caption{Dis-equilibrium chemical composition for the atmosphere of HD 189733 b assuming the \citet{Showman09} day-average GCM temperature profile (dashed lines) and the dis-equilibrium composition temperature profile with soot haze calculated in this work for complete (4$\pi$) redistribution (solid lines).}
\label{fig:HD189_comp}
\end{figure}

\begin{figure}
\includegraphics[width=\columnwidth]{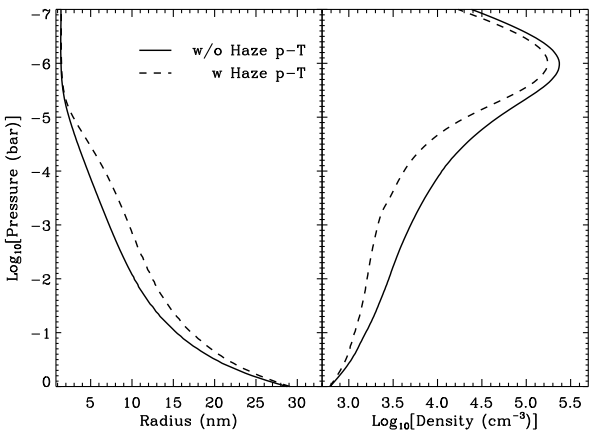}
\caption{Average particle radius and corresponding particle density in the atmosphere of HD 189733 b for two temperature profiles corresponding to the cases with and without the impact of soot type photochemical hazes. The particle mass flux is set at 10$^{-11}$ gcm$^{-2}$s$^{-1}$.}\label{fig:HD189_haze}
\end{figure}

\begin{figure*}
\includegraphics[scale=0.35]{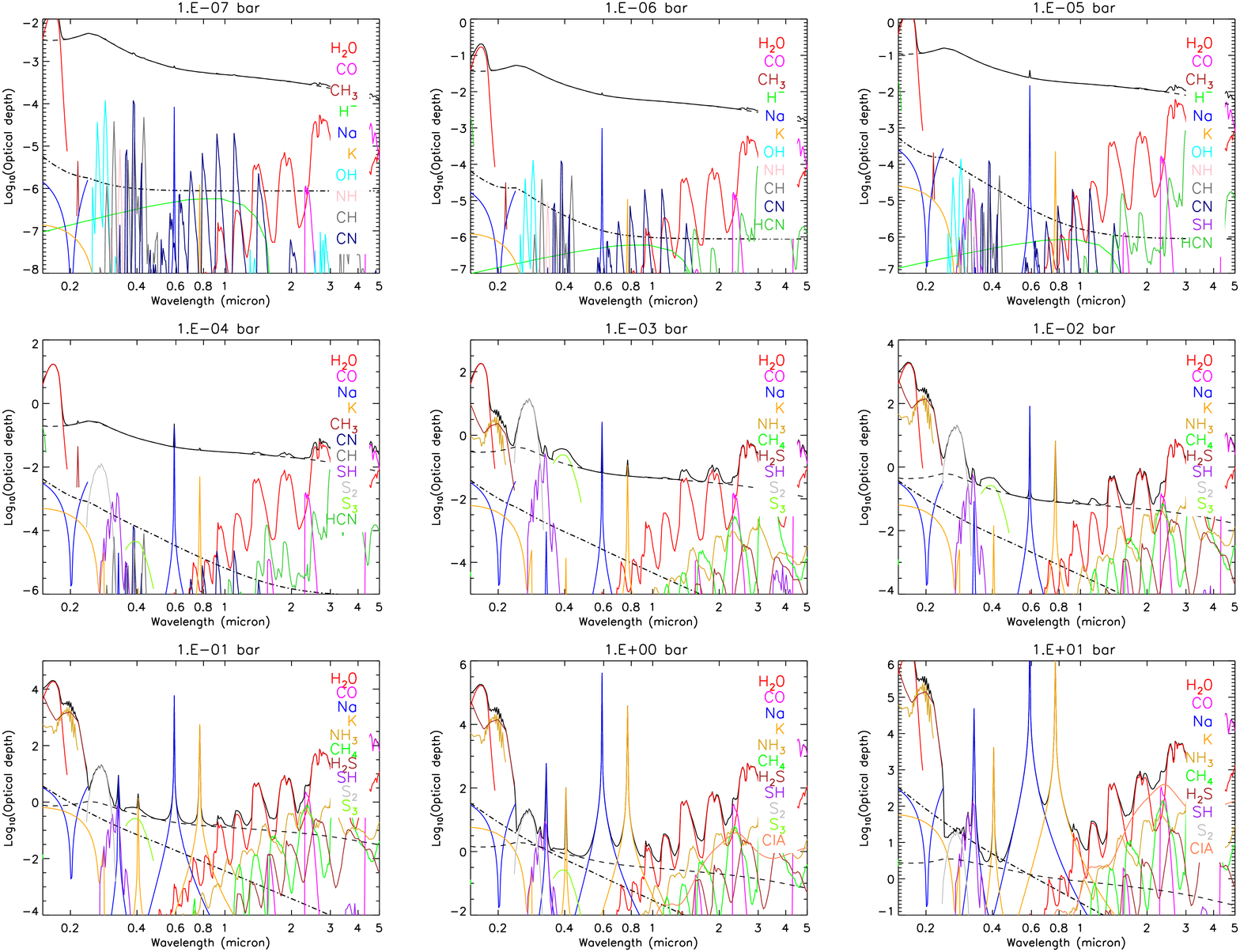}
\caption{Gas and photochemical haze opacity contributions across wavelength for stellar energy deposition calculations at different pressures in the atmosphere of HD 189733 b. Different colours present the absorption optical depth contribution by different gases based on the dis-equilibrium chemical composition. The dashed black line presents the absorption optical depth by the haze particles and the solid black line the total absorption optical depth (gas and haze). The dash-dotted lines present the total scattering optical depth.}
\label{fig:opacity_gas_HD189}
\end{figure*}

\subsection{HD 189733 b}

\begin{figure}
\includegraphics[width=\columnwidth]{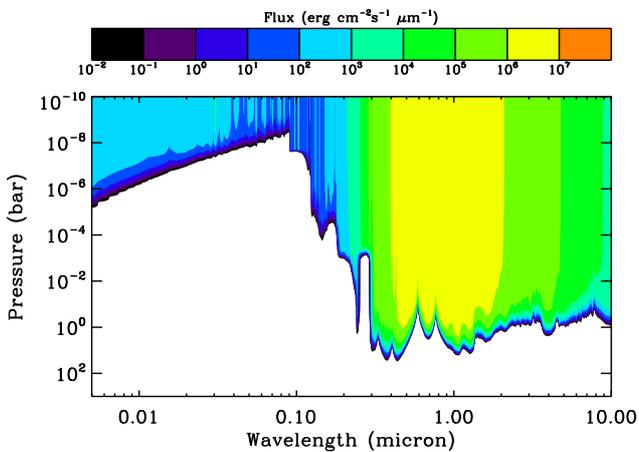}
\caption{Stellar flux penetration in the atmosphere of HD 189733 b at different wavelengths. }
\label{fig:flx_str_189}
\end{figure}

\begin{figure*}
\includegraphics[width=\columnwidth]{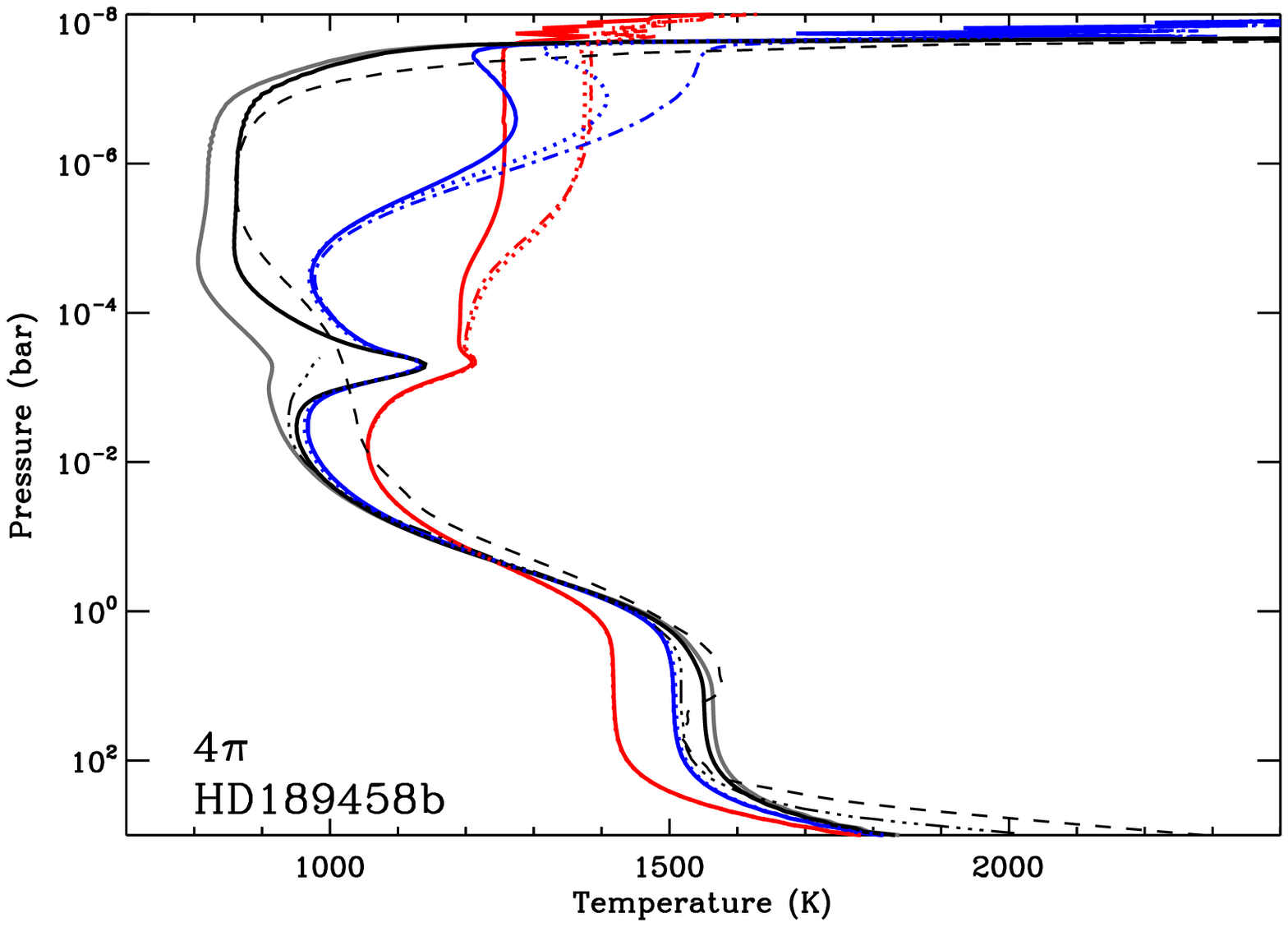}
\includegraphics[width=\columnwidth]{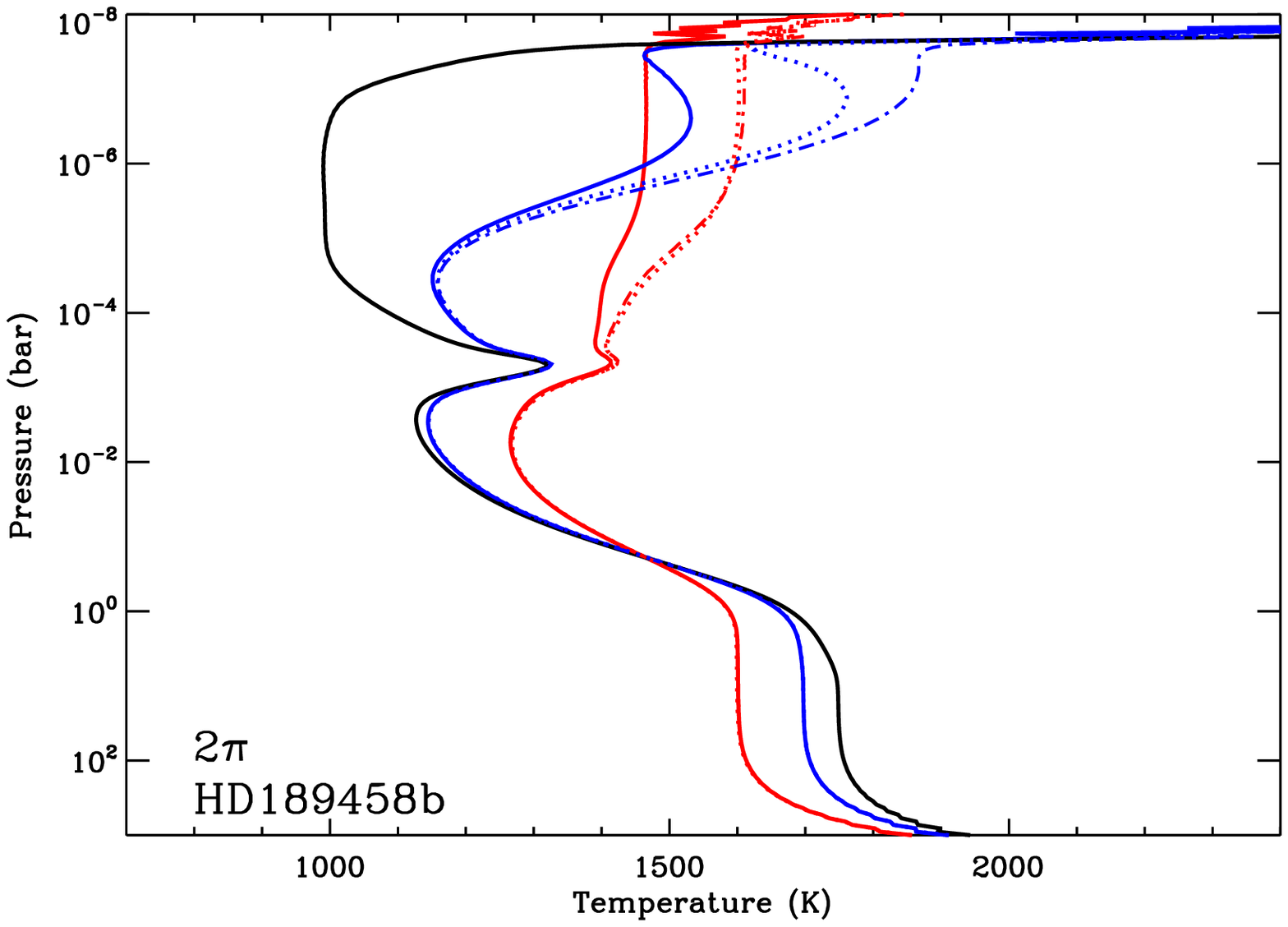}
\caption{Temperature profiles for HD 189733 b for 4$\pi$ (left) and 2$\pi$ (right) energy redistribution. Solid lines present our simulated profiles of disequilibrium chemistry without hazes (black), with soot opacity haze (red), and with Titan-type opacity haze (blue). The grey line presents the case of no haze and without the opacity from sulfur species (see text). Previous studies from 1D \citep[dash-dotted,][]{Fortney06} and 3D \citep[dashed,][]{Showman09} simulations are also presented, with the latter corresponding to the temperature profile used for the evaluation of the dis-equilibrium chemical composition.  The dash-dotted lines present the average particle temperature demonstrating the disequilibrium between particles and gas, while the dotted lines present the resulting atmospheric temperature if the disequilibrium between particles and gas is not taken into account. For all simulations convection is limited to the first layer near the lower boundary (10$^{3}$ bar).}
\label{fig:temp189}
\end{figure*}

HD 189733 b is considered a paradigm of a hazy hot-Jupiter. For the evaluation of its thermal structure we consider the disequilibrium chemical composition and photochemical haze size distribution derived previously \citep{Lavvas17}. These chemical composition results were based on a general circulation temperature profile and eddy mixing coefficient \citep{Showman09} assuming thermochemical equilibrium (with condensation, i.e. no TiO and VO) and solar elemental abundances. The photochemical haze particle size distribution was derived based on the observed transit depth and a detailed microphysics model. The input chemical composition and the temperature profile used for its evaluation are shown with dashed lines in Figs.~\ref{fig:HD189_comp} \& \ref{fig:temp189}, respectively. The particle average properties are shown in Fig.~\ref{fig:HD189_haze}. HD 189733 is a K star and we consider for its spectrum a composite of  $\epsilon$-Eridani and a PHOENIX model at the stellar parameters \citep{Lavvas17}. As the qualitative results for this atmosphere are similar to that of HD 209458 b, we focus here on the impact of disequilibrium chemistry and photochemical hazes.

The gaseous opacity distribution of HD 189733 b is similar to that of HD 209458 b presented earlier. However, for HD 189733 b more molecules contribute substantially to the total opacity such as NH$_3$ (near 0.2$\mu$m) and $S_3$ (near 0.4$\mu$m), while HCN and CH$_4$ contribute in the H$_2$O bands troughs in the near-IR (Fig.~\ref{fig:opacity_gas_HD189}). This difference is a result of the survival of H$_2$O to higher altitudes in this atmosphere, allowing other species such as CH$_4$ and NH$_3$ to survive (through UV screening and through reduction of atomic hydrogen abundance), and contribute to the atmospheric opacity. Moreover, the photolysis of the latter species leads to a higher population of primary radicals such as OH, CH, NH, SH and CN  in the upper atmosphere of HD 189733 b compared to that of HD 209458 b, with a corresponding signature on the opacity spectrum.

Nevertheless, the major difference of HD 189733 b compared to the opacity of HD 209458 b is the photochemical haze contribution. The latter is the dominant opacity  source in the upper atmosphere for pressure levels above 0.1 mbar, while it significantly contributes to the visible opacity between the atomic/molecular lines down to $\sim$1 bar (Fig.~\ref{fig:opacity_gas_HD189}). This evaluation is based on the soot refractive index. However, assuming the Titan-type refractive index does not modify qualitatively the picture; although the total opacity reduces, haze remains the dominant absorber in the upper atmosphere.

Considering only the gaseous composition, our simulated profile for complete redistribution (4$\pi$) and T$_{int}$ = 100 K is similar to previous evaluations of thermochemical equilibrium (black lines in Fig.~\ref{fig:temp189}). Specifically, our calculations are in good agreement in the lower atmosphere below 10 mbar, with equilibrium composition 1D simulations \citep{Fortney06}, as well as with the day-average profile from the \cite{Showman09} general circulation model. However, as observed in the case of HD 209458 b, the disequilibrium contributions of SH, S$_2$ (and also S$_3$ for HD 189733 b) provide a significant opacity that drives a local temperature increase centred at 0.5 mbar, which was not present in previous studies. At the peak, the temperature difference with the case where SH, S$_2$ and S$_3$ opacities are not included (grey line in Fig.~\ref{fig:temp189}) is $\sim$220 K. Moreover, it is important to highlight that the difference in the abundances of these species between equilibrium and disequilibrium evaluation is much larger than in the case of HD 209458 b, and results in a more clear signature in the penetrating stellar fluxes in the atmosphere of HD 189733 b near 0.3 $\mu$m (Fig.~\ref{fig:flx_str_189}). Therefore, in this cooler planet consideration of disequilibrium processes has a pronounced impact on the thermal structure. 

At pressures above 0.1$\mu$bar the lack of efficient atmospheric cooling due to the destruction of H$_2$O results in a rapid temperature increase as identified in hydrodynamic escape models \citep{Koskinen13}.

Inclusion of the haze (soot) opacity in the thermal structure simulation results in major heating of the upper atmosphere at pressures lower than 0.1 bar and with the maximum temperature increase reaching $\sim$400 K at 1$\mu$bar (solid red line in Fig.~\ref{fig:temp189}). As more photons are absorbed in the upper atmosphere, heating of the lower atmosphere is reduced resulting in a temperature drop of $\sim$130 K in the isothermal part of the atmosphere near 10 bar. Assuming the Titan-type refractive index, the corresponding temperature changes are still significant (solid blue line in Fig.~\ref{fig:temp189}). In this case the presence of haze particles leads to a similar temperature increase of $\sim$400 K near 1 $\mu$bar, but the vertical extend of the temperature rise is smaller and the profile merges with the clear atmosphere profile below $\sim$1 mbar. This is a direct consequence of the lower absorptivity of Titan-type material relative to soot that allows more of the visible/near-IR radiation to penetrate to deeper atmospheric layers. As a result the corresponding cooling of the lower atmosphere is limited to $\sim$45 K relative to the clear atmosphere case. Nevertheless, within the range of optical properties considered here, we find that photochemical hazes will have a significant impact on the atmospheric thermal structure. 

We must highlight at this point that we considered the same particle size distribution for both refractive index temperature evaluations. This size distribution derived from the microphysics model was constrained by fitting the observed transit spectrum with a particle mass flux of 5$\times$10$^{-12}$g cm$^{-2}$s$^{-1}$ and considering the soot refractive index and a temperature profile that did not include the photochemical haze impact \citep{Lavvas17}. As discussed below, assuming the Titan-type refractive index with the same size distribution results in a transit spectrum that falls short of the observed, therefore a higher particle mass flux would need to be considered for this type of particle composition. The required higher opacity will increase further the role of such particles in heating the upper atmosphere, as identified above for the soot composition.

\begin{figure}
\includegraphics[width=\columnwidth]{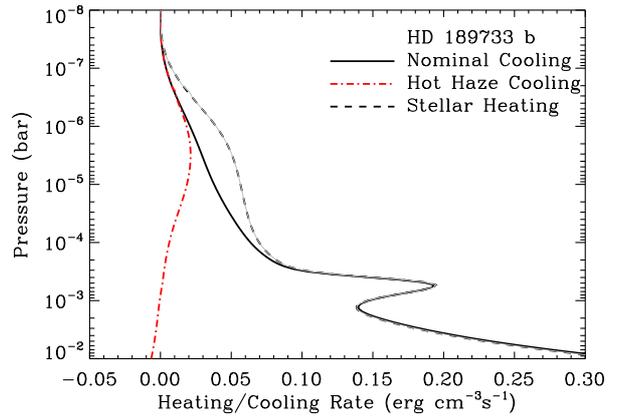}
\caption{Stellar heating (dashed line) and thermal cooling rates at the atmospheric temperature (solid line) and the excess emission from soot type hot-particles (red line). The thin grey line overlapping the dashed line is the sum of thermal rates, thus demonstrates the energy balance in the calculations.}
\label{fig:cooling}
\end{figure}

The strong interaction of stellar radiation with the haze particles leads to their rapid heating (Fig.~\ref{fig:temp189}). For the soot composition assumption, the average particle temperature diverges from the gas temperature at p < 0.1 mbar and remains $\sim$100 K higher than the atmospheric temperature (with haze heating included) at pressure levels above 1$\mu$bar. For the Titan-type composition particles become hotter than the gas above $\sim$10$\mu$bar and reach a maximum temperature difference of $\sim$300K at lower pressures. Although soot composition particles absorb more efficiently, their resulting temperature is lower from those assuming Titan-type composition. This feature relates to the higher atmospheric density of the upper atmosphere for the soot case, which results in a more efficient energy transfer from particles to gas, relative to the Titan-type case. Note that at all cases the temperature difference between the particles and the gas remains small enough (T$_p$/T$_g$<1.3) for the description of energy exchange rate to be accurate. 

The strong difference between gas and particle temperatures demonstrates that the inclusion of the excess particle thermal emission is important for the correct evaluation of the energy balance in the upper atmosphere. For the soot composition, particle emission peaks between 0.1 $\mu$bar and 1 $\mu$bar but its global impact to the atmospheric energy balance is important above 0.1 mbar (Fig.~\ref{fig:cooling}). For Titan-type particles the effect is reduced but the picture is qualitatively the same. If particles were assumed to be in equilibrium with the gas the resulting atmospheric temperature near 1~$\mu$bar would be higher by $\sim$100-200 K depending on the particle refractive index assumed (see dotted lines in Fig.~\ref{fig:temp189}).

\begin{figure}
\includegraphics[width=\columnwidth]{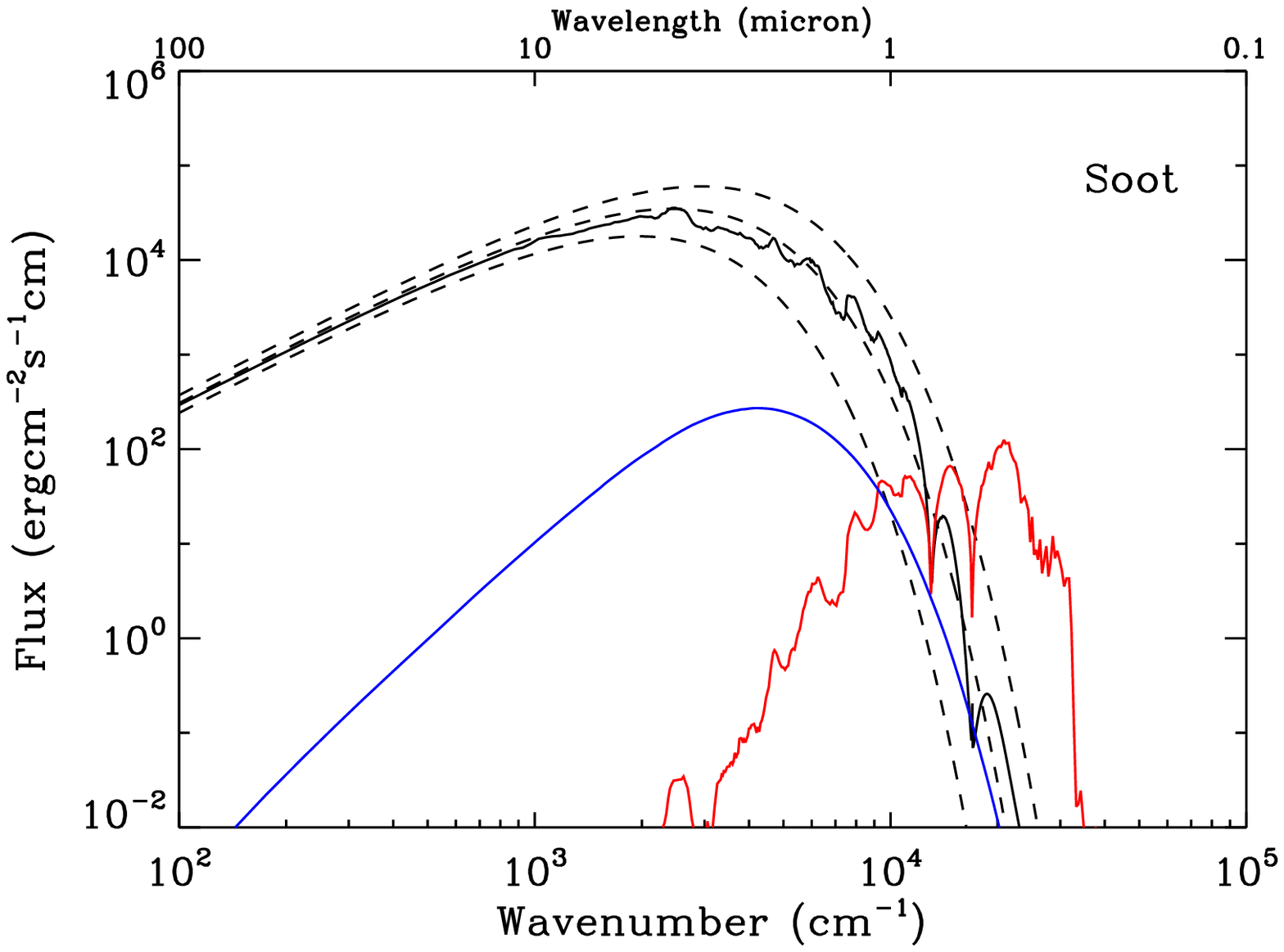}
\includegraphics[width=\columnwidth]{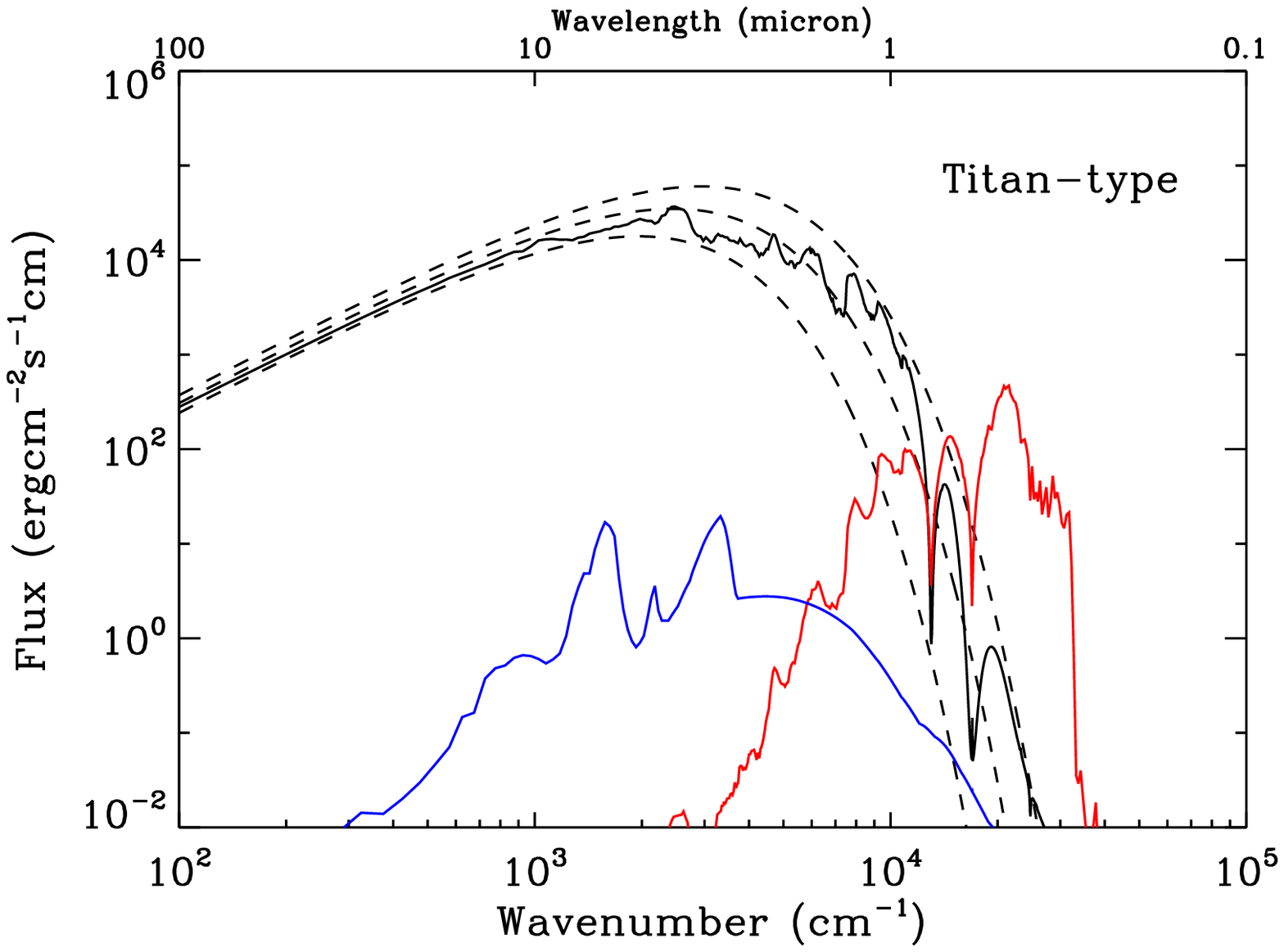}
\caption{Planetary thermal emission from gases (black) and particles (blue) and stellar reflected radiation (red) at the top of the simulated atmosphere, assuming 4$\pi$-redistribution. Dashed lines present black body emission at 1000 K, 1250 K and 1500 K. The top panel corresponds to soot composition particles and the bottom to Titan-type composition.}
\label{fig:emission189}
\end{figure}

The radiation emitted by the hot particles reflects the refractive index assumed. The high absorptivity of soots over the whole spectrum results in a black-body emission spectrum, while for Titan-type particles the strong absorption features in the near IR leave a strong signature to the emitted spectrum (Fig.~\ref{fig:emission189}). However, at the top of the atmosphere the particle emission is negligible compared to the gas emission from deeper layers, thus it would not be observable. As with the case of HD 209458 b, the emitted radiation dominantly originates from levels between 1 bar and 0.1 mbar. At the top of the atmosphere thermally emitted radiation dominates at wavelengths longer than the K line and reflected stellar radiation prevails at shorter wavelengths.
 
\section{Implications}

Our results demonstrate that the impact of photochemical hazes and disequilibrium chemistry on the atmospheric temperature can be significant. These effects could have further ramifications for the interpretation of transit and secondary eclipse observations. Thus, we explore these implications here.

\begin{figure*}
\includegraphics[scale=0.42]{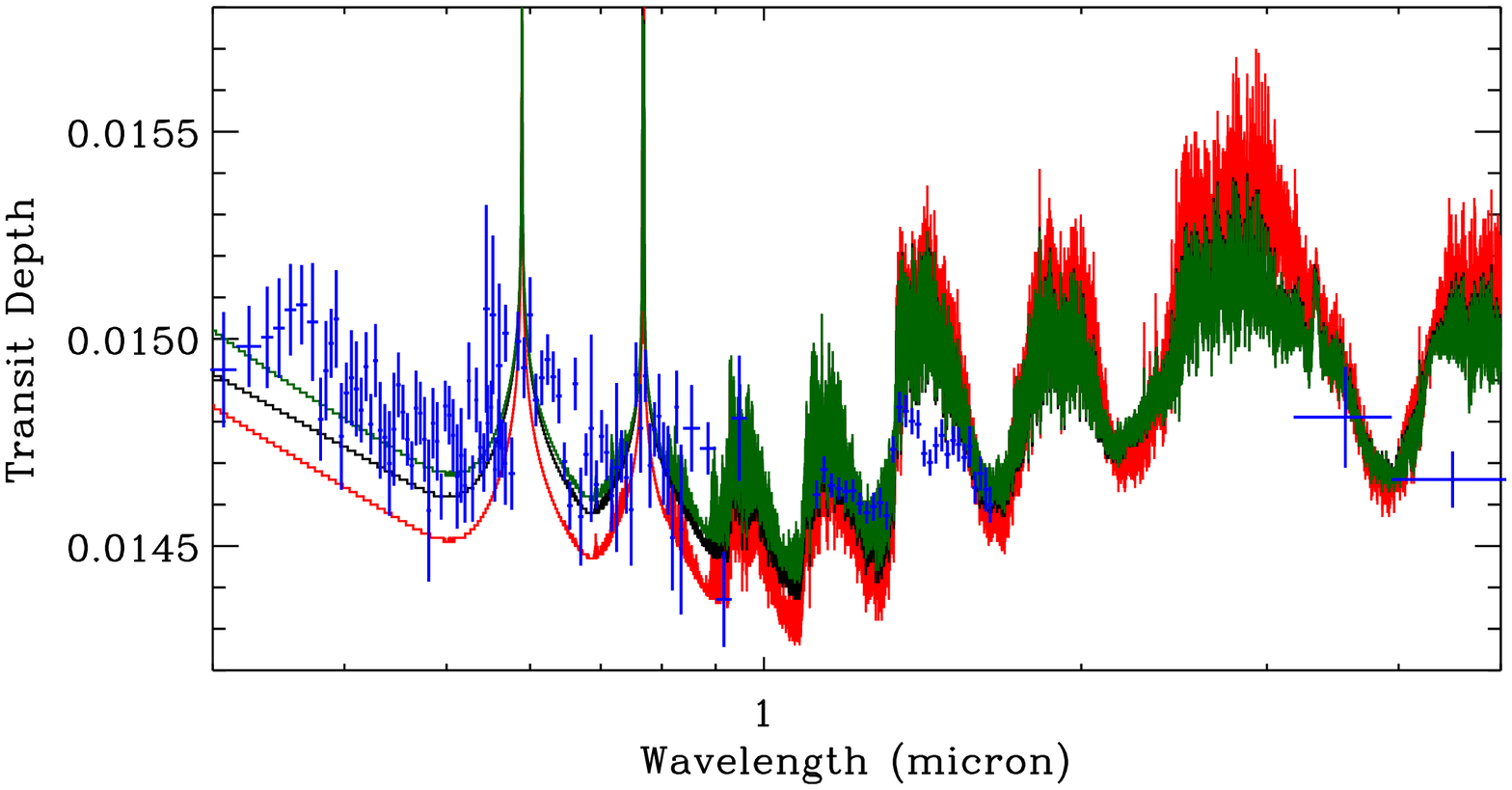}
\includegraphics[scale=0.4]{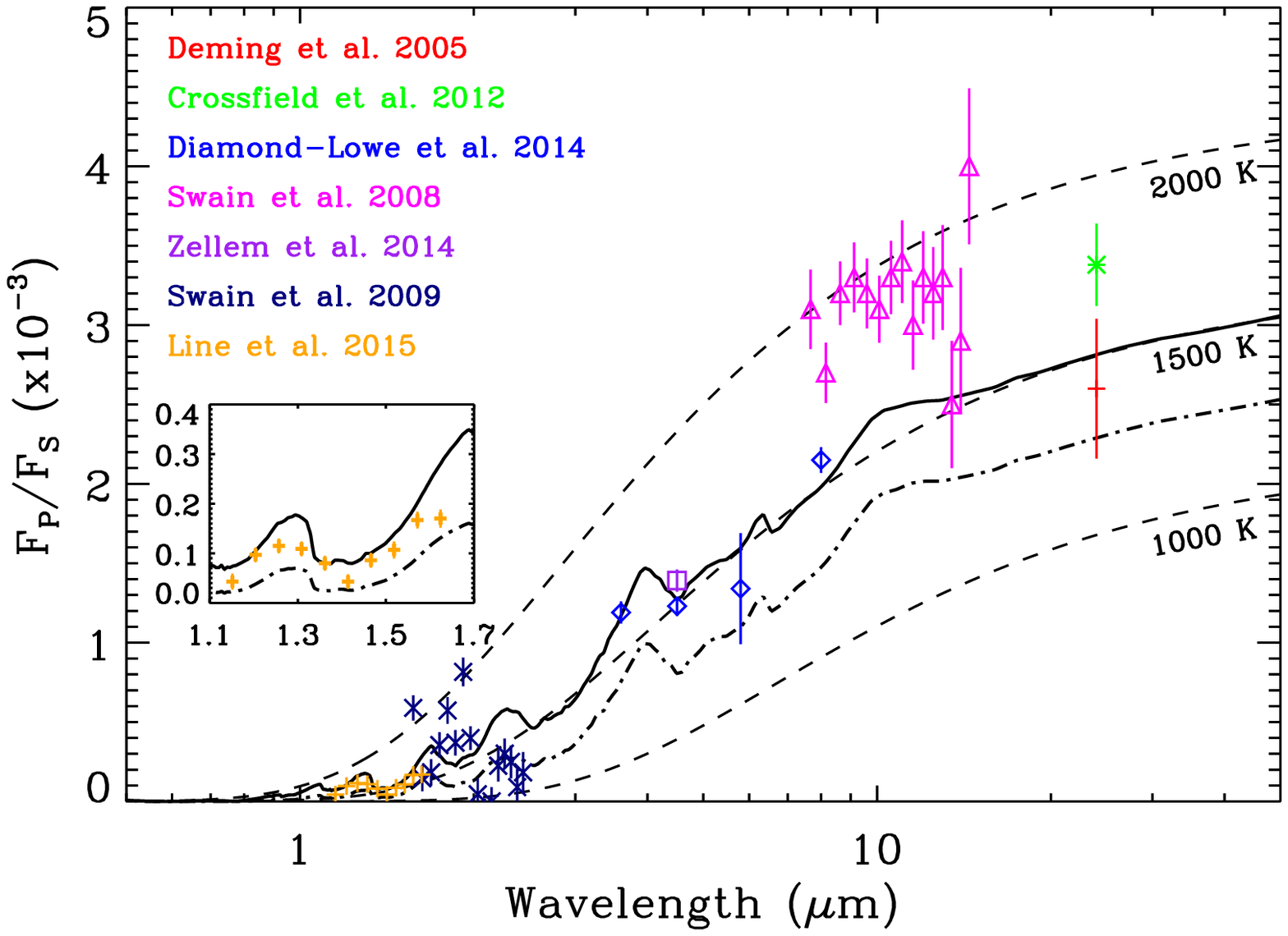}
\caption{Observed (symbols) and simulated (lines) transit (left) and secondary eclipse (right) spectra of HD 209458 b. Transit observations are from \citet{Sing16}. The red solid line presents the simulated spectrum assuming the equilibrium composition temperature profile (i.e. utilises the chemistry and temperature profiles presented by dashed lines in Figs.~\ref{fig:HD209_comp} and \ref{fig:temp209}), the black solid line assumes the disequilibrium composition temperature profile (i.e utilises the chemistry profiles presented by solid lines in Fig.~\ref{fig:HD209_comp} and black line temperature profile in Fig.~\ref{fig:temp209}), and the green solid line assumes the disequilibrium + photochemical haze profile discussed in the text. For the eclipse spectrum observations are from \citet{Deming05} (red cross), \citet{Swain08} (cyan triangles), \citet{Swain09} (navy asterisks), \citet{Crossfield12} (green asterisk), \citet{Diamond14} (blue diamonds), \citet{Zellem14} (purple square), and \citet{Line16} (orange crosses, see also inset). Dashed lines present theoretical planet/star flux ratios for planetary black body emissions at the presented temperatures. The solid line is our simulated spectrum for the disequilibrium chemical composition assuming energy deposition on the day side, while the dash-dotted line spectrum assumes complete redistribution.}
\label{fig:transit209}
\end{figure*}
\subsection{HD 209458 b}

Our previous evaluation of the transit signature of HD 209458 b suggested a rather clear atmosphere, at least for the part of the atmosphere probed by the observations \citep{Lavvas14}. However that conclusion was based on the GCM temperature profile that included the thermochemical equilibrium abundances of TiO and VO, i.e. corresponding to a significantly warmer atmosphere than the one we calculate here. This was actually a critical factor in suggesting that the higher temperature of HD 209458 b relative to that of HD 189733 b was responsible for the lack of photochemical hazes in the former atmosphere: the photochemical composition simulations demonstrated that the main gas species considered as precursors of the photochemical hazes were significantly reduced at high temperatures. The cooler temperature profile under the disequilibrium chemical composition questions this conclusion, thus necessitates a re-evaluation of the chemical composition and photochemistry mass fluxes. 

The observational constraints for the atmosphere of HD 209458 b also demand a re-evaluation of the presence of hazes. A new analysis of HST/STIS observations \citep{Sing16} revealed the presence of the missing potassium \citep{Lavvas14}, but also resulted in a significantly stronger transit signature in the UV-Visible part of the spectrum relative to the previous analyses \citep{Sing08,Deming13}. Therefore, the updated constraints may be consistent with the presence of hazes in this atmosphere, after all \citep{Pinhas19}. 

To address these concerns we re-evaluated the dis-equilibrium chemical composition considering the new temperature profile evaluated in this study, all other parameters (e.g. stellar flux, eddy mixing profile) remaining the same as in our previous evaluation \citep{Lavvas14}. We consider the complete redistribution profile for this evaluation as the disk average conditions it corresponds to, are likely more representative of the atmospheric limbs probed in transit. Clearly the cooler temperature affects the chemical composition of the upper atmosphere with most prominent modification the transition from H$_2$ dominance to H dominance moving from 0.1 mbar in the hot temperature profile to $\sim$1 $\mu$bar in the cooler temperature profile (Fig.~\ref{fig:HD209_comp}). This modification has further ramifications for other species that are affected by the abundance of atomic hydrogen such as H$_2$O, N$_2$ and CO for which their mixing ratios increase in the upper atmosphere. Using the revised thermal structure and chemical composition, the simulated transit spectrum (black line in Fig.~\ref{fig:transit209}) demonstrates a stronger signature in the UV-Visible slope compared to the spectrum based on the equilibrium thermal structure (red line). Most importantly, the revised observations demonstrate an even higher transit depth in the same spectral region, suggesting that a haze contribution could be possible.

The main photochemical haze precursors such as CH$_4$, HCN and C$_2$H$_2$ demonstrate higher abundances at the disequilibrium temperature. The most significant increase appears in the profile of HCN that attains a mixing ratio of $\sim$10$^{-6}$ in the upper atmosphere, while the profiles of CH$_4$ and C$_2$H$_2$ also increase but their mixing ratios remain in the order of 10$^{-9}$. The corresponding mass fluxes from the photolysis of these species reflects the changes in their abundances (Table~\ref{tab:mass_flux_HD209}). The mass flux generated by the photolysis of HCN at pressure levels above 10$\mu$bar is $\sim$60$\times$ larger from our previous evaluation based on the equilibrium temperature profile \citep{Lavvas14}, providing a mass flux of 4.8$\times$10$^{-13}$ gcm$^{-2}$s$^{-1}$. Similarly, photolysis of CH$_4$ and C$_2$H$_2$ generate mass fluxes that are $\sim$10$^{6}\times$ and $\sim$10$^{3}\times$ larger than previous estimates, respectively. Previously, we estimated the mass flux for soot particles from the photolysis of these three species considering a moderate yield of 1 per cent. Such an estimate for the new conditions investigated results in a mass flux of 5.2$\times$10$^{-15}$ gcm$^{-2}$s$^{-1}$, which is $\sim$70$\times$ larger than the previous evaluation. Our previous conclusion regarding the role of high temperature as a modulator of the photochemical haze production remains valid: the revised mass flux in the atmosphere of HD 209458 b is still significantly smaller than that in the cooler atmosphere of HD 189733 b. Nevertheless, the updated mass flux magnitude may suggest that photochemical hazes could be present in the atmosphere of HD 209458 b. 

\begin{table}
\caption{Photolysis mass fluxes (in gcm$^{-2}$s$^{-1}$) of major species in the upper atmosphere (p$<$10$\mu$bar) of HD 209458 b and HD 189733 b under the two temperature profiles of equilibrium (E) and disequilibrium (DE) composition of each case. The soot mass flux is estimated from the sum of CH$_4$, C$_2$H$_2$ and HCN mass fluxes assuming a moderate 1$\%$ efficiency \citep{Lavvas17}.}\label{tab:mass_flux_HD209}
\centering
\begin{tabular}{lcccc}
\hline
	&\multicolumn{2}{c}{HD 209458 b}  & \multicolumn{2}{c}{HD 189733 b}  \\
\hline
Species & E p-T & DE p-T& E p-T & DE p-T \\
\hline
CH$_4$ 		& 1.1(-22) & 6.2(-16) & 7.1(-16) & 2.8(-16) \\
HCN 		& 7.8(-15) & 4.8(-13)	& 2.9(-11) & 2.1(-10) \\
C$_2$H$_2$ 	& 2.8(-17) & 4.1(-14) & 1.8(-11) & 8.6(-12) \\
CO 			& 4.8(-11) & 5.9(-12)	& 4.8(-12) & 5.4(-13) \\
N$_2$ 		&  4.5(-12) & 5.1(-13)& 7.0(-13)& 2.7(-12)\\
NH$_3$ 		& 3.5(-16) & 2.9(-13)	& 1.6(-15)& 3.0(-15)\\
H$_2$S 		& 1.8(-11) & 1.5(-11)	& 5.9(-13)& 6.3(-13)\\
S$_3$ 		& 1.8(-19) & 2.0(-12)	& 1.4(-13)& 3.2(-14)\\
\hline
Soot (1$\%$)& 7.8(-17) & 5.2(-15) & 4.7(-13) & 2.2(-12)\\
\hline
\end{tabular}
\end{table}

We therefore considered how a photochemical haze mass flux with the above magnitude would affect the transit spectrum. Our preliminary evaluation of the particle size distribution suggests that the effect would be negligible. However, for this evaluation we assumed the eddy mixing profile derived from the atmospheric circulation simulations that considered the equilibrium chemical composition with TiO/VO \citep{Showman09}, thus is not representative of the conditions investigated here. Assuming a weaker eddy mixing, as also suggested by subsequent evaluations with GCM models \citep{Parmentier13, Zhang18}, allows for a more efficient particle growth \citep{Lavvas17}. A 100$\times$ weaker eddy profile results in the green transit spectrum in Fig.~\ref{fig:transit209} that agrees better to the observed UV-Visible transit depth. However, a higher formation yield for photochemical hazes is also possible \citep{Lavvas17}. Moreover, photolysis of CO and CO$_2$ may further contribute to the mass flux of photochemical hazes, as recent laboratory experiments suggest, albeit of a smaller yield relative to the pathways based on methane \citep{He18}. 

On the other hand, the revised observations indicate the presence of other features in the transit spectrum. Particularly features at the Na wings and at shorter wavelengths are reminiscent of gaseous opacities that would suggest there are missing components in the gaseous abundances considered. In the past we evaluated the contributions of heavier atoms such as Al, Ca, Mg, Fe and Si in the transit spectrum \citep[see fig.10 in][]{Lavvas14}, assuming that these are not lost due to condensation, with the results suggesting that they could contribute significantly to the opacity. However, such contributions are expected to be far more significant in ultra hot exoplanets \citep{Lothringer20}. On the contrary, the low temperature conditions we calculate here suggest that heavy metals should remain condensed in the atmosphere of HD 209458 b and should not contribute to the atmospheric opacity. This conclusion is further supported by a recent re-analysis of the NUV spectrum of HD 209458 b that did not detect the presence of Mg/Mg$^+$ in the upper atmosphere \citep{Cubillos20}, although the detection of Fe beyond the planetary roche lobe is a concern.
Moreover, the observed transit shape of H$_2$O at 1.4 $\mu$m appears also affected by the haze, which is not reproduced by the few nm radius particles resulting from our microphysics simulations. This characteristic further indicates that condensate clouds could also be affecting the transit depth at near IR wavelengths. Based on our transit simulation the pressures probed at wavelengths $\lambda$$<$2$\mu$m are larger than 1 mbar, which makes the possibility of photochemical haze interaction with cloud formation very likely (see condensation curves for silicates in Fig.~\ref{fig:temp209}). All these uncertainties demonstrate the complexity of the problem, as well as, the need for further investigation of the interaction mechanisms among different types of particles and their feedback to the gas background.

\begin{figure*}
\includegraphics[scale=0.43]{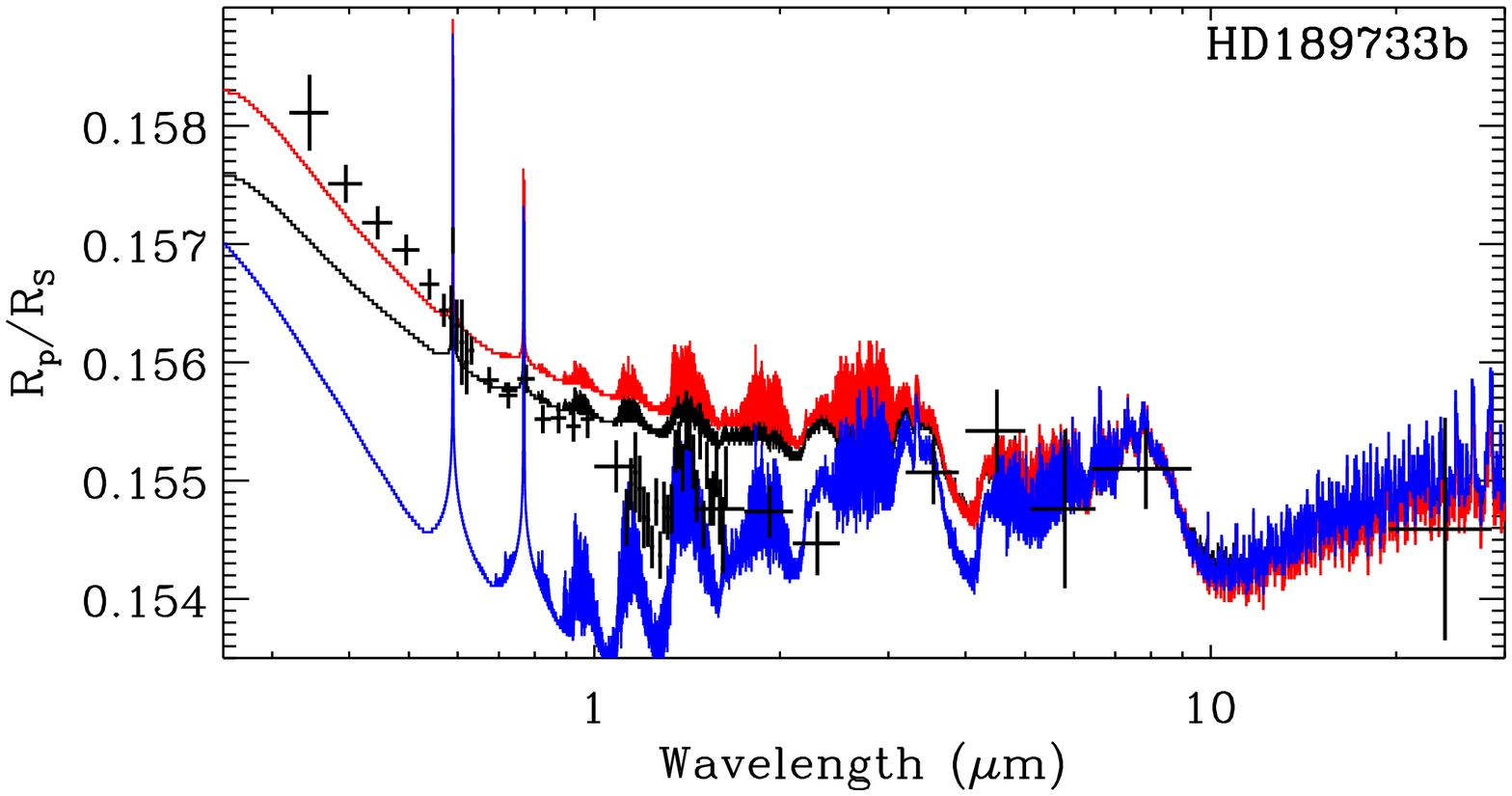}
\includegraphics[scale=0.4]{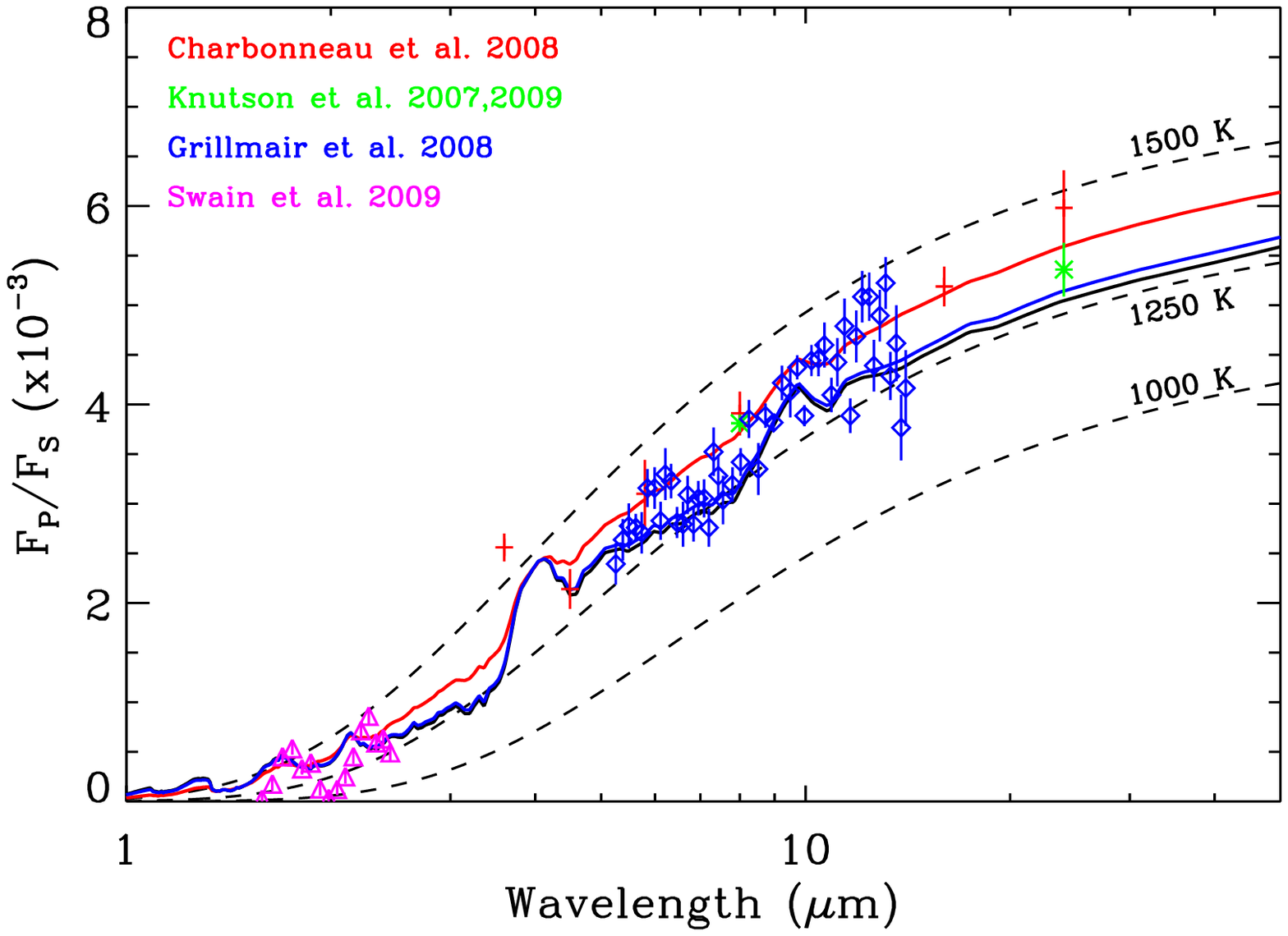}
\caption{Observed (symbols) and simulated (lines) transit (left) and secondary eclipse (right) spectra of HD 189733 b. For the transit spectrum observations are from \citet{Sing16}. Lines present simulations assuming the thermochemical equilibrium temperature profile (black) and the disequilibrium temperature profiles evaluated here under the soot (red) and the Titan-type (blue) composition assumptions (for all cases the chemical composition is described by the dashed profiles in Fig.~\ref{fig:HD189_comp}). For the secondary eclipse spectrum observations are from \citet{Charbonneau08} (red crosses), \citet{Knutson07,Knutson09} (green asterisks), \citet{Grillmair08} (blue diamonds), \citet{Swain09} (cyan triangles). Dashed lines are theoretical planet/star flux ratios for planetary black body emissions at the presented temperatures. Solid lines present the simulated spectrum for the cases presented earlier (black: no haze, red: soot haze, blue: Titan-type haze) evaluated assuming energy deposition on the day side.}
\label{fig:transit189}
\end{figure*}

We need to keep in mind that there are further complications arising from the 3D aspect of the atmosphere. These complications are better demonstrated by the comparison with the secondary eclipse observations.
Our simulations assuming complete redistribution fall short of the observations, but assuming energy deposition only on the day side provides a spectrum that is in better agreement (Fig.~\ref{fig:transit209}). The corresponding temperature profile (green line in Fig.~\ref{fig:temp209}) is $\sim$200-300 K hotter than the nominal case of complete redistribution, and consistent with the \citet{Line16}  temperature retrievals based on secondary eclipse observations (see inset in Fig.~\ref{fig:temp209}). However, even with the 2$\pi$-redistribution the simulated eclipse spectrum does demonstrate a lower emission near 10 $\mu$m relative to the observations. Our calculations suggest that photons in this spectral region mainly originate from pressure levels between 1 mbar and 1 bar, i.e. where also our transit simulations suggest that possible hazes should reside. The presence of photochemical hazes could locally heat the atmosphere and increase the thermally emitted radiation, bringing the simulated emission spectrum in better agreement with the observations, as demonstrated below for the case of HD 189733 b. The same effect could also result from the presence of clouds. Thus, the secondary eclipse observations may further support the presence of aerosols in the atmosphere of HD 209458 b.

\subsection{HD 189733 b}
For HD 189733 b, we note that the transit at short wavelengths ($\lambda$$\lesssim$1$\mu$m) probes the upper atmosphere at pressures between $\sim$1 mbar and $\sim$1 $\mu$bar \citep{Lavvas17}. As demonstrated above, the inclusion of photochemical haze opacity affects this part of the atmosphere and the involved changes to the atmospheric scale height could modify the simulated transit spectra. Therefore, we evaluated the difference in the transit spectrum of HD 189733 b assuming the temperature profiles with and without haze (Fig.~\ref{fig:transit189}). The results reveal that the hotter upper atmosphere with the inclusion of soot haze is significantly more extended yielding a stronger signature to the transit depth by increasing the UV-Visible slope (red line). The impact of hazes on the thermal structure, may therefore reduce the requirement for a strong eddy mixing to keep the particles aloft at high altitudes \citep{Ohno20} and potentially allow for a convergence between the eddy mixing profile necessary to explain observations and those estimated by GCM models \citep{Parmentier13, Zhang18}.

On the other hand, considering the Titan-type composition demonstrates the requirement for a re-evaluation of the appropriate mass flux and size distribution; the resulting transit depth assuming the same particle size distribution as for soot particles falls short of the observations (blue line) due to the lower opacity demonstrated by the Titan-type composition. A preliminary evaluation suggests that a factor of $\sim$10 higher opacity would be required to match the transit spectrum at visible and UV wavelengths assuming a Titan-type refractive index. Such an increase in opacity would also bring the temperature closer to the profile derived for the soot composition particles. Thus, the important conclusion from this exercise is that the correct interpretation of transit spectra for hazy exoplanets requires proper consideration of the haze impact on the thermal structure. 

Our temperature profiles presented above furthermore demonstrate that inclusion of photochemical hazes also affects the photosphere therefore could modify the secondary eclipse signature. Our simulated eclipse spectrum for HD 189733 b is consistent with the observations for all simulated cases assuming energy depositions on the day side (Fig.~\ref{fig:transit189}). Nevertheless, the temperature changes imposed by the inclusion of haze opacity reduces the difference between model and observation (particularly for $\lambda$>10$\mu$m), with the impact being more pronounced for the soot composition haze that has a stronger influence to the part of the atmosphere affecting the thermally emitted radiation (see Fig.~\ref{fig:temp189}). Although this trend is clear, a definite conclusion regarding the role of photochemical hazes on the eclipse spectrum requires further investigation; photochemical hazes could act as nucleation sites for the formation of clouds in the photosphere, which will modify both the size distribution and optical properties of the resulting particles. Therefore a complete evaluation of the impact of hazes on the atmospheric structure and the interpretation of eclipse observations requires a coupled photochemical haze-cloud simulation. With the advent of more accurate observations in the forthcoming years through JWST, ARIEL, and other observatories, it will be possible to set more accurate constraints on the contribution of clouds and hazes.

The temperature increase in the upper atmosphere of HD 189733 b imposed by the photochemical haze opacity will have an impact on both the atmospheric chemistry and the formation and evolution of the haze particles, which we have not yet taken into account. Therefore, our work here presents only part of the implications from the inclusion of such particles.  Considering only gas phase chemistry \cite{Drummond16} presented self-consistent simulations of disequilibrium chemical composition and temperature and found that the coupling reduced the departure from equilibrium chemistry composition. However, the haze opacity in our simulations is orders of magnitude higher than the gaseous opacity at visible wavelengths (Fig.~\ref{fig:opacity_gas_HD189}). This dramatic difference in opacity sources demonstrates why the potential changes in the chemical composition from the self-consistent description are not likely to modify our conclusion on the impact of hazes on the atmospheric thermal structure.

As a demonstration of the anticipated ramifications, our disequilibrium chemical composition simulations using the disequilibrium temperature profile with soot haze opacity reveal non-negligible changes in the profiles of the main gaseous species (Fig.~\ref{fig:HD189_comp}). These changes reflect modifications of the chemical kinetic rates and diffusion properties of the species through their temperature dependence. Note that temperature also affects the thermochemical equilibrium that controls the lower boundary conditions for the main species. Thus, results in further modifications of the main abundances depending on the elemental composition and T$_{int}$ value assumed \citep{Fortney20}. The disequilibrium temperature profile results in higher abundances for H$_2$O, CH$_4$ and NH$_3$ in the photosphere region, while the H$_2$ to H transition starts at lower pressures near 0.1 mbar. CH$_4$ and C$_2$H$_2$ have smaller abundances in the upper atmosphere, compared to the profiles under the equilibrium temperature profile, while HCN  demonstrates a strong increase in the whole atmosphere. The mass fluxes from the photolysis of the main species are indeed different from the previous evaluation based on the equilibrium composition temperature profile (Table~\ref{tab:mass_flux_HD209}) suggesting a higher production of photochemical hazes.

However, even without the changes in the background gas composition, the higher temperature in the upper atmosphere will independently affect the microphysical evolution of the hazes.
As a demonstration of the impact of temperature on the evolution of the haze particles we compared their microphysical evolution under the clear atmosphere and soot-composition haze scenarios (Fig.~\ref{fig:HD189_haze}). The results show that the higher temperature drives a more efficient collision rate of the particles that yields a faster particle growth in the upper atmosphere relative to the case of the clear atmosphere temperature profile. Note that the temperature increase is not sufficient to sublimate the particles, as seen in the lower atmosphere \citep{Lavvas17}, therefore does not pose a problem for the presence of the particles at the simulated conditions. 

Finally, we can note from the simulated transit spectra (Fig.~\ref{fig:transit189}) that a refractive index similar to the soot, but with a lower absorptivity towards the near IR would be more consistent with the observations. However, a re-evaluation of the microphysics taking into account the impact on the thermal structure is required before any definite conclusions. The revised chemical composition and haze distribution will further affect the thermal structure necessitating a new temperature evaluation as well as a new evaluation of the optimum haze mass fluxes to match the transit spectrum. Therefore it is clear that a self-consistent description of thermal structure, atmospheric chemistry and haze microphysics is necessary for a proper evaluation of the feedback among each domain, and the interpretation of current and future observations.

\section{Conclusions}

We describe a detailed radiative-convective model for the investigation of the atmospheric thermal structure of hot exoplanets and investigate the impact of photochemical hazes. We find that the inclusion of such particles result in major heating of the upper and cooling of the lower atmosphere. The magnitude of this effect depends on the refractive index assumed for the optical properties of the particles, but our simulations demonstrate that the effect is important for the highly absorbing soot composition, as well as, for the relatively more scattering Titan-type composition. 

We also considered the role of disequilibrium chemical composition relative to the usual assumption of thermochemical equilibrium (with or without rainout) in previous thermal structure evaluations. We find that sulphur containing species, such as SH, S$_2$ and S$_3$ provide significant opacity in the middle atmosphere and lead to local heating near 1 mbar. Similarly OH, CH, NH, and CN radicals produced by the photochemistry affect the thermal structure near 1$\mu$bar.

Furthermore we show that the modifications on the thermal structure from photochemical gases and hazes can have important ramifications for the interpretation of transit observations. Our study for the hazy HD 189733 b shows that the hotter upper atmosphere resulting from the inclusion of photochemical haze opacity imposes an expansion of the atmosphere, thus a steeper transit signature in the UV-Visible part of the spectrum. In addition, the temperature changes in the photosphere also affect the secondary eclipse spectrum. For HD 209458 b we find that the cooler atmospheric thermal structure under disequilibrium chemical composition provides a transit spectrum that is no longer consistent with the latest observational constraints. This conclusions suggests that some haze opacity could be present in this atmosphere, at pressures below 1 mbar, which could be a result of both photochemical hazes and condensates. However further investigation is required with a self-consistent coupled description.

The above conclusions for the role of hazes are particularly important for general circulation models that investigate the horizontal distribution of particles by the winds and are able to evaluate the whole phase curve variation of the planetary emission. Currently most studies that involve particles follow a tracer approach that reveals how fixed size particles are distributed in the atmosphere  but do not include the feedback of particle absorption to the thermal structure and the winds field. Our study motivates the inclusion of this feedback in future GCM simulations.

\section*{Acknowledgements}

We acknowledge support by the Programme National de Plan\'etologie through the project TISSAGE. We thank R.V. Yelle for providing the S$_2$ linelist and T.T. Koskinen for comments on the manuscript. We also thank our reviewer for constructive comments and suggestions that resulted in an improved manuscript.
%

\section*{Data availability}

The data underlying this article will be shared on reasonable request to the corresponding author.




\bibliographystyle{mnras}
\bibliography{refs} 

\begin{thebibliography}{}
\makeatletter
\relax
\def\mn@urlcharsother{\let\do\@makeother \do\$\do\&\do\#\do\^\do\_\do\%\do\~}
\def\mn@doi{\begingroup\mn@urlcharsother \@ifnextchar [ {\mn@doi@}
  {\mn@doi@[]}}
\def\mn@doi@[#1]#2{\def\@tempa{#1}\ifx\@tempa\@empty \href
  {http://dx.doi.org/#2} {doi:#2}\else \href {http://dx.doi.org/#2} {#1}\fi
  \endgroup}
\def\mn@eprint#1#2{\mn@eprint@#1:#2::\@nil}
\def\mn@eprint@arXiv#1{\href {http://arxiv.org/abs/#1} {{\tt arXiv:#1}}}
\def\mn@eprint@dblp#1{\href {http://dblp.uni-trier.de/rec/bibtex/#1.xml}
  {dblp:#1}}
\def\mn@eprint@#1:#2:#3:#4\@nil{\def\@tempa {#1}\def\@tempb {#2}\def\@tempc
  {#3}\ifx \@tempc \@empty \let \@tempc \@tempb \let \@tempb \@tempa \fi \ifx
  \@tempb \@empty \def\@tempb {arXiv}\fi \@ifundefined
  {mn@eprint@\@tempb}{\@tempb:\@tempc}{\expandafter \expandafter \csname
  mn@eprint@\@tempb\endcsname \expandafter{\@tempc}}}

\bibitem[\protect\citeauthoryear{Adams, Gao, de Pater  \& Morley}{Adams
  et~al.}{2019}]{Adams19}
Adams D.,  Gao P.,  de Pater I.,   Morley C.~V.,  2019, The Astrophysical
  Journal, 874:61, 14 pp

\bibitem[\protect\citeauthoryear{Amaral, Diniz, Jones, Stanke, Alijah,
  Adamowicz  \& Mohallem}{Amaral et~al.}{2019}]{Amaral19}
Amaral P. H.~R.,  Diniz L.~G.,  Jones K.~A.,  Stanke M.,  Alijah A.,  Adamowicz
  L.,   Mohallem J.~R.,  2019, The Astrophysical Journal, 878:95, 14 pp

\bibitem[\protect\citeauthoryear{Amundsen, Baraffe, Tremblin, Manners, Hayek,
  Mayne  \& Acreman}{Amundsen et~al.}{2014}]{Amundsen14}
Amundsen D.~S.,  Baraffe I.,  Tremblin P.,  Manners J.,  Hayek W.,  Mayne
  N.~J.,   Acreman D.~M.,  2014, Astronomy and Astrophysics, 564, A59

\bibitem[\protect\citeauthoryear{Arney et~al.,}{Arney et~al.}{2016}]{Arney16}
Arney G.,  et~al., 2016, Astrobiology, 16, 873

\bibitem[\protect\citeauthoryear{Azzam, Tennyson, Yurchenko  \& Naumenko}{Azzam
  et~al.}{2016}]{Azzam16}
Azzam A. A.~A.,  Tennyson J.,  Yurchenko S.~N.,   Naumenko O.~V.,  2016,
  Monthly Notices of the Royal Astronomical Society, 460, 4063

\bibitem[\protect\citeauthoryear{Barber, Strange, Hill, Polyansky, Mellau,
  Yurchenko  \& Tennyson}{Barber et~al.}{2013}]{Barber13}
Barber R.~J.,  Strange J.~K.,  Hill C.,  Polyansky O.~L.,  Mellau G.~C.,
  Yurchenko S.~N.,   Tennyson J.,  2013, Monthly Notices of the Royal
  Astronomical Society, 437, 1828

\bibitem[\protect\citeauthoryear{Bell}{Bell}{1980}]{Bell80}
Bell K.~L.,  1980, Journal of Physics B: Atomic and Molecular Physics, 13, 1859

\bibitem[\protect\citeauthoryear{Bell \& Berrington}{Bell \&
  Berrington}{1987}]{Bell87}
Bell K.~L.,  Berrington K.~A.,  1987, Journal of Physics B: Atomic and
  Molecular Physics, 20, 801

\bibitem[\protect\citeauthoryear{Brooke, Ram, Western, Li, Schwenke  \&
  Bernath}{Brooke et~al.}{2014}]{Brooke14}
Brooke J. S.~A.,  Ram R.~S.,  Western C.~M.,  Li G.,  Schwenke D.~W.,   Bernath
  P.~F.,  2014, The Astrophysical Journal Supplement Series, 210, 23

\bibitem[\protect\citeauthoryear{Burrows \& Sharp}{Burrows \&
  Sharp}{1999}]{Burrows99}
Burrows A.,  Sharp C.,  1999, Astrophysical Journal, 512, 843

\bibitem[\protect\citeauthoryear{Burrows et~al.,}{Burrows
  et~al.}{1997}]{Burrows97}
Burrows A.,  et~al., 1997, Astrophysical Journal v.491, 491, 856

\bibitem[\protect\citeauthoryear{Chandrasekhar}{Chandrasekhar}{1960}]{Chandrasekhar}
Chandrasekhar S.,  1960, Radiative Transfer.
Dover

\bibitem[\protect\citeauthoryear{Charbonneau, Knutson, Barman, Allen, Mayor,
  Megeath, Queloz  \& Udry}{Charbonneau et~al.}{2008}]{Charbonneau08}
Charbonneau D.,  Knutson H.~A.,  Barman T.,  Allen L.~E.,  Mayor M.,  Megeath
  S.~T.,  Queloz D.,   Udry S.,  2008, The Astrophysical Journal, 686, 1341

\bibitem[\protect\citeauthoryear{Charnay, Meadows, Misra, Leconte  \&
  Arney}{Charnay et~al.}{2015}]{Charnay15}
Charnay B.,  Meadows V.,  Misra A.,  Leconte J.,   Arney G.,  2015, The
  Astrophysical Journal Letters, 813:L1, 7 pp

\bibitem[\protect\citeauthoryear{Chubb, Tennyson  \& Yurchenko}{Chubb
  et~al.}{2020}]{Chubb20}
Chubb K.~L.,  Tennyson J.,   Yurchenko S.~N.,  2020, Monthly Notices of the
  Royal Astronomical Society, 493, 1531

\bibitem[\protect\citeauthoryear{Coles, Yurchenko  \& Tennyson}{Coles
  et~al.}{2019}]{Coles19}
Coles P.~A.,  Yurchenko S.~N.,   Tennyson J.,  2019, Monthly Notices of the
  Royal Astronomical Society, 490, 4638

\bibitem[\protect\citeauthoryear{Crossfield \& Kreidberg}{Crossfield \&
  Kreidberg}{2017}]{Crossfield17}
Crossfield I. J.~M.,  Kreidberg L.,  2017, The Astronomical Journal, 154:261, 6
  pp

\bibitem[\protect\citeauthoryear{Crossfield, Knutson, Fortney, Showman, Cowan
  \& Deming}{Crossfield et~al.}{2012}]{Crossfield12}
Crossfield I. J.~M.,  Knutson H.,  Fortney J.,  Showman A.~P.,  Cowan N.~B.,
  Deming D.,  2012, The Astrophysical Journal, 752:81, 13 pp

\bibitem[\protect\citeauthoryear{Cubillos, Fossati, Koskinen, Young, Salz,
  France, Sreejith  \& Haswell}{Cubillos et~al.}{2020}]{Cubillos20}
Cubillos P.~E.,  Fossati L.,  Koskinen T.,  Young M.~E.,  Salz M.,  France K.,
  Sreejith A.~G.,   Haswell C.~A.,  2020, The Astronomical Journal, 159, 111

\bibitem[\protect\citeauthoryear{Deming, Seager, Richardson  \&
  Harrington}{Deming et~al.}{2005}]{Deming05}
Deming D.,  Seager S.,  Richardson L.~J.,   Harrington J.,  2005, Nature, 434,
  740

\bibitem[\protect\citeauthoryear{Deming et~al.,}{Deming
  et~al.}{2013}]{Deming13}
Deming D.,  et~al., 2013, The Astrophysical Journal, 774:95, 17 pp

\bibitem[\protect\citeauthoryear{Diamond-Lowe, Stevenson, Bean, Line  \&
  Fortney}{Diamond-Lowe et~al.}{2014}]{Diamond14}
Diamond-Lowe H.,  Stevenson K.~B.,  Bean J.~L.,  Line M.~R.,   Fortney J.~J.,
  2014, The Astrophysical Journal, 796:66, 7 pp

\bibitem[\protect\citeauthoryear{Drummond, Tremblin, Baraffe, Amundsen, Mayne,
  Venot  \& Goyal}{Drummond et~al.}{2016}]{Drummond16}
Drummond B.,  Tremblin P.,  Baraffe I.,  Amundsen D.~S.,  Mayne N.~J.,  Venot
  O.,   Goyal J.,  2016, Astronomy and Astrophysics, 594, A69

\bibitem[\protect\citeauthoryear{Fernando, Bernath, Hodges  \&
  Masseron}{Fernando et~al.}{2018}]{Fernando18}
Fernando A.~M.,  Bernath P.~F.,  Hodges J.~N.,   Masseron T.,  2018, Journal of
  Quantitative Spectroscopy and Radiative Transfer, 217, 29

\bibitem[\protect\citeauthoryear{Filippov \& Rosner}{Filippov \&
  Rosner}{1999}]{Filippov99}
Filippov A.~V.,  Rosner D.~E.,  1999, International Journal of Heat and Mass
  Transfer, 43, 127

\bibitem[\protect\citeauthoryear{Fleury, Gudipati, Henderson  \& Swain}{Fleury
  et~al.}{2019}]{Fleury19}
Fleury B.,  Gudipati M.~S.,  Henderson B.~L.,   Swain M.,  2019, The
  Astrophysical Journal, 871:158, 14 pp

\bibitem[\protect\citeauthoryear{Fortney, Sudarsky, Hubeny, Cooper, Hubbard,
  Burrows  \& Lunine}{Fortney et~al.}{2003}]{Fortney03}
Fortney J.~J.,  Sudarsky D.,  Hubeny I.,  Cooper C.~S.,  Hubbard W.~B.,
  Burrows A.,   Lunine J.~I.,  2003, The Astrophysical Journal, 589, 615

\bibitem[\protect\citeauthoryear{Fortney, Marley, Lodders, Saumon  \&
  Freedman}{Fortney et~al.}{2005}]{Fortney05}
Fortney J.~J.,  Marley M.~S.,  Lodders K.,  Saumon D.,   Freedman R.,  2005,
  The Astrophysical Journal, 627, L69

\bibitem[\protect\citeauthoryear{Fortney, Saumon, Marley, Lodders  \&
  Freedman}{Fortney et~al.}{2006}]{Fortney06}
Fortney J.~J.,  Saumon D.,  Marley M.~S.,  Lodders K.,   Freedman R.~S.,  2006,
  The Astrophysical Journal, 642, 495

\bibitem[\protect\citeauthoryear{Fortney, Lodders, Marley  \& Freedman}{Fortney
  et~al.}{2008}]{Fortney08}
Fortney J.~J.,  Lodders K.,  Marley M.~S.,   Freedman R.~S.,  2008, The
  Astrophysical Journal, 678, 1419

\bibitem[\protect\citeauthoryear{Fortney, Visscher, Marley, Hood, Line,
  Thorngren, Freedman  \& Lupu}{Fortney et~al.}{2020}]{Fortney20}
Fortney J.~J.,  Visscher C.,  Marley M.~S.,  Hood C.~E.,  Line M.~R.,
  Thorngren D.~P.,  Freedman R.~S.,   Lupu R.,  2020, Beyond Equilibrium
  Temperature: How the Atmosphere/Interior Connection Affects the Onset of
  Methane, Ammonia, and Clouds in Warm Transiting Giant Planets (\mn@eprint
  {arXiv} {2010.00146})

\bibitem[\protect\citeauthoryear{Gao \& Benneke}{Gao \& Benneke}{2018}]{Gao18}
Gao P.,  Benneke B.,  2018, The Astrophysical Journal, 863:165, 23 pp

\bibitem[\protect\citeauthoryear{Gao, Marley, Zahnle, Robinson  \& Lewis}{Gao
  et~al.}{2017}]{Gao17}
Gao P.,  Marley M.~S.,  Zahnle K.,  Robinson T.~D.,   Lewis N.~K.,  2017, The
  Astronomical Journal, 153, 139

\bibitem[\protect\citeauthoryear{Gao et~al.,}{Gao et~al.}{2020}]{Gao20}
Gao P.,  et~al., 2020, Nature Astronomy, 4, 951

\bibitem[\protect\citeauthoryear{Gombosi}{Gombosi}{1994}]{Gombosi}
Gombosi T.,  1994, Gaskinetic Theory.
Cambridge

\bibitem[\protect\citeauthoryear{Goody \& {Yung}}{Goody \&
  {Yung}}{1989}]{Goody}
Goody R.,  {Yung} Y.,  1989, Atmospheric Radiation.
Oxford

\bibitem[\protect\citeauthoryear{Gorman, Yurchenko  \& Tennyson}{Gorman
  et~al.}{2019}]{Gorman19}
Gorman M.~N.,  Yurchenko S.~N.,   Tennyson J.,  2019, Monthly Notices of the
  Royal Astronomical Society, 490, 1652

\bibitem[\protect\citeauthoryear{Grillmair et~al.,}{Grillmair
  et~al.}{2008}]{Grillmair08}
Grillmair C.~J.,  et~al., 2008, Nature, 456, 767

\bibitem[\protect\citeauthoryear{Guillot \& Showman}{Guillot \&
  Showman}{2002}]{Guillot02}
Guillot T.,  Showman A.~P.,  2002, Astronomy and Astrophysics, 385, 156

\bibitem[\protect\citeauthoryear{Guillot, Chabrier, Morel  \& Gautier}{Guillot
  et~al.}{1994}]{Guillot94}
Guillot T.,  Chabrier G.,  Morel P.,   Gautier D.,  1994, Icarus (ISSN
  0019-1035), 112, 354

\bibitem[\protect\citeauthoryear{He et~al.,}{He et~al.}{2018}]{He18}
He C.,  et~al., 2018, The Astronomical Journal, 156:38, 8 pp

\bibitem[\protect\citeauthoryear{He et~al.,}{He et~al.}{2020}]{He20}
He C.,  et~al., 2020, The Planetary Science Journal, 1, 1

\bibitem[\protect\citeauthoryear{Heijden \& Mullen}{Heijden \&
  Mullen}{2001}]{Heijden01}
Heijden H. v.~d.,  Mullen J. v.~d.,  2001, Journal Of Physics B-Atomic
  Molecular And Optical Physics, 34, 4183

\bibitem[\protect\citeauthoryear{Helling et~al.,}{Helling
  et~al.}{2019}]{Helling19}
Helling C.,  et~al., 2019, Astronomy and Astrophysics, 631, A79

\bibitem[\protect\citeauthoryear{Iro, Bezard  \& Guillot}{Iro
  et~al.}{2005}]{Iro05}
Iro N.,  Bezard B.,   Guillot T.,  2005, Astronomy and Astrophysics, 436, 719

\bibitem[\protect\citeauthoryear{Israel et~al.,}{Israel
  et~al.}{2005}]{Israel05}
Israel G.,  et~al., 2005, Nature, 438, 796

\bibitem[\protect\citeauthoryear{John}{John}{1988}]{John88}
John T.~L.,  1988, Astronomy and Astrophysics (ISSN 0004-6361), 193, 189

\bibitem[\protect\citeauthoryear{Kawashima \& Ikoma}{Kawashima \&
  Ikoma}{2018}]{Kawashima18}
Kawashima Y.,  Ikoma M.,  2018, The Astrophysical Journal, 853:7, 26 pp

\bibitem[\protect\citeauthoryear{Kawashima, Hu  \& Ikoma}{Kawashima
  et~al.}{2019}]{Kawashima19}
Kawashima Y.,  Hu R.,   Ikoma M.,  2019, The Astrophysical Journal Letters,
  876, L5

\bibitem[\protect\citeauthoryear{Kempton, Bean  \& Parmentier}{Kempton
  et~al.}{2017}]{Kempton17}
Kempton E. M.~R.,  Bean J.~L.,   Parmentier V.,  2017, The Astrophysical
  Journal Letters, 845, L20

\bibitem[\protect\citeauthoryear{Knutson et~al.,}{Knutson
  et~al.}{2007}]{Knutson07}
Knutson H.~A.,  et~al., 2007, Nature, 447, 183

\bibitem[\protect\citeauthoryear{Knutson et~al.,}{Knutson
  et~al.}{2009}]{Knutson09}
Knutson H.~A.,  et~al., 2009, The Astrophysical Journal, 690, 822

\bibitem[\protect\citeauthoryear{Komacek, Showman  \& Parmentier}{Komacek
  et~al.}{2019}]{Komacek19}
Komacek T.~D.,  Showman A.~P.,   Parmentier V.,  2019, The Astrophysical
  Journal, 881:152, 23 pp

\bibitem[\protect\citeauthoryear{Koskinen, Harris, Yelle  \& Lavvas}{Koskinen
  et~al.}{2013}]{Koskinen13}
Koskinen T.~T.,  Harris M.~J.,  Yelle R.~V.,   Lavvas P.,  2013, Icarus, 226,
  1678

\bibitem[\protect\citeauthoryear{Lacis \& Oinas}{Lacis \&
  Oinas}{1991}]{Lacis91}
Lacis A.~A.,  Oinas V.,  1991, Journal of Geophysical Research, 96, 9027

\bibitem[\protect\citeauthoryear{Lavvas \& Koskinen}{Lavvas \&
  Koskinen}{2017}]{Lavvas17}
Lavvas P.,  Koskinen T.,  2017, The Astrophysical Journal, 847:32, 20 pp

\bibitem[\protect\citeauthoryear{Lavvas, Coustenis  \& Vardavas}{Lavvas
  et~al.}{2008}]{Lavvas08}
Lavvas P.~P.,  Coustenis A.,   Vardavas I.~M.,  2008, Planetary And Space
  Science, 56, 27

\bibitem[\protect\citeauthoryear{Lavvas, Koskinen  \& Yelle}{Lavvas
  et~al.}{2014}]{Lavvas14}
Lavvas P.,  Koskinen T.,   Yelle R.~V.,  2014, The Astrophysical Journal,
  796:15, 20 pp

\bibitem[\protect\citeauthoryear{Lavvas, Koskinen, Steinrueck, Mu{\~n}oz  \&
  Showman}{Lavvas et~al.}{2019}]{Lavvas19}
Lavvas P.,  Koskinen T.,  Steinrueck M.~E.,  Mu{\~n}oz A.~G.,   Showman A.~P.,
  2019, The Astrophysical Journal, 878:118, 16 pp

\bibitem[\protect\citeauthoryear{Li, Gordon, Rothman, Tan, Hu, Kassi, Campargue
   \& Medvedev}{Li et~al.}{2015}]{Li15}
Li G.,  Gordon I.~E.,  Rothman L.~S.,  Tan Y.,  Hu S.-M.,  Kassi S.,  Campargue
  A.,   Medvedev E.~S.,  2015, The Astrophysical Journal Supplement Series,
  216:15, 18 pp

\bibitem[\protect\citeauthoryear{Line et~al.,}{Line et~al.}{2016}]{Line16}
Line M.~R.,  et~al., 2016, The Astronomical Journal, 152, 203

\bibitem[\protect\citeauthoryear{Lines et~al.,}{Lines et~al.}{2018}]{Lines18}
Lines S.,  et~al., 2018, Monthly Notices of the Royal Astronomical Society,
  481, 194

\bibitem[\protect\citeauthoryear{Lothringer, Fu, Sing  \& Barman}{Lothringer
  et~al.}{2020}]{Lothringer20}
Lothringer J.~D.,  Fu G.,  Sing D.~K.,   Barman T.~S.,  2020, The Astrophysical
  Journal Letters, 898, L14

\bibitem[\protect\citeauthoryear{Masseron et~al.,}{Masseron
  et~al.}{2014}]{Masseron14}
Masseron T.,  et~al., 2014, {AstronomyAstrophysics}, 571:A47, 29 pp

\bibitem[\protect\citeauthoryear{Mbarek \& Kempton}{Mbarek \&
  Kempton}{2016}]{Mbarek16}
Mbarek R.,  Kempton E. M.~R.,  2016, The Astrophysical Journal, 827, 1

\bibitem[\protect\citeauthoryear{McKay, Pollack  \& Courtin}{McKay
  et~al.}{1989}]{McKay89}
McKay C.~P.,  Pollack J.~B.,   Courtin R.,  1989, Icarus (ISSN 0019-1035), 80,
  23

\bibitem[\protect\citeauthoryear{McKay, Pollack  \& Courtin}{McKay
  et~al.}{1991}]{McKay91}
McKay C.~P.,  Pollack J.~B.,   Courtin R.,  1991, Science (ISSN 0036-8075),
  253, 1118

\bibitem[\protect\citeauthoryear{McKemmish, Yurchenko  \& Tennyson}{McKemmish
  et~al.}{2016}]{McKemmish16}
McKemmish L.~K.,  Yurchenko S.~N.,   Tennyson J.,  2016, Monthly Notices of the
  Royal Astronomical Society, 463, 771

\bibitem[\protect\citeauthoryear{Mizus, Alijah, Zobov, Lodi, Kyuberis,
  Yurchenko, Tennyson  \& Polyansky}{Mizus et~al.}{2017}]{Mizus17}
Mizus I.~I.,  Alijah A.,  Zobov N.~F.,  Lodi L.,  Kyuberis A.~A.,  Yurchenko
  S.~N.,  Tennyson J.,   Polyansky O.~L.,  2017, Monthly Notices of the Royal
  Astronomical Society, 468, 1717

\bibitem[\protect\citeauthoryear{Morisson, Szopa, Carrasco, Buch  \&
  Gautier}{Morisson et~al.}{2016}]{Morisson16}
Morisson M.,  Szopa C.,  Carrasco N.,  Buch A.,   Gautier T.,  2016, Icarus,
  277, 442

\bibitem[\protect\citeauthoryear{Morley, Fortney, Kempton, Marley, Visscher  \&
  Zahnle}{Morley et~al.}{2013}]{Morley13}
Morley C.~V.,  Fortney J.~J.,  Kempton E. M.~R.,  Marley M.~S.,  Visscher C.,
  Zahnle K.,  2013, The Astrophysical Journal, 775, 33

\bibitem[\protect\citeauthoryear{Morley, Fortney, Marley, Zahnle, Line,
  Kempton, Lewis  \& Cahoy}{Morley et~al.}{2015}]{Morley15}
Morley C.~V.,  Fortney J.~J.,  Marley M.~S.,  Zahnle K.,  Line M.,  Kempton E.,
   Lewis N.,   Cahoy K.,  2015, The Astrophysical Journal, 815, 110

\bibitem[\protect\citeauthoryear{Moses et~al.,}{Moses et~al.}{2011}]{Moses11}
Moses J.~I.,  et~al., 2011, The Astrophysical Journal, 737:15, 37 pp

\bibitem[\protect\citeauthoryear{Ohno \& Kawashima}{Ohno \&
  Kawashima}{2020}]{Ohno20}
Ohno K.,  Kawashima Y.,  2020, The Astrophysical Journal Letters, 895: L47, 9
  pp

\bibitem[\protect\citeauthoryear{Ohno \& Okuzumi}{Ohno \&
  Okuzumi}{2018}]{Ohno18}
Ohno K.,  Okuzumi S.,  2018, The Astrophysical Journal, 859:34, 17 pp

\bibitem[\protect\citeauthoryear{Parmentier, Showman  \& Lian}{Parmentier
  et~al.}{2013}]{Parmentier13}
Parmentier V.,  Showman A.~P.,   Lian Y.,  2013, Astronomy and Astrophysics,
  558, A91

\bibitem[\protect\citeauthoryear{Pinhas, Madhusudhan, Gandhi  \&
  MacDonald}{Pinhas et~al.}{2019}]{Pinhas19}
Pinhas A.,  Madhusudhan N.,  Gandhi S.,   MacDonald R.,  2019, Monthly Notices
  of the Royal Astronomical Society, 482, 1485

\bibitem[\protect\citeauthoryear{Plez}{Plez}{1998}]{Plez98}
Plez B.,  1998, Astronomy and Astrophysics, 337, 495

\bibitem[\protect\citeauthoryear{Polyansky, Kyuberis, Zobov, Tennyson,
  Yurchenko  \& Lodi}{Polyansky et~al.}{2018}]{Polyansky18}
Polyansky O.~L.,  Kyuberis A.~A.,  Zobov N.~F.,  Tennyson J.,  Yurchenko S.~N.,
    Lodi L.,  2018, Monthly Notices of the Royal Astronomical Society, 480,
  2597

\bibitem[\protect\citeauthoryear{Powell, Zhang, Gao  \& Parmentier}{Powell
  et~al.}{2018}]{Powell18}
Powell D.,  Zhang X.,  Gao P.,   Parmentier V.,  2018, The Astrophysical
  Journal, 860:18, 26 pp

\bibitem[\protect\citeauthoryear{Pruppacher \& Klett}{Pruppacher \&
  Klett}{1997}]{Pruppacher}
Pruppacher H.,  Klett J.,  1997, Microphysics of clouds and precipitation.
Kluwer Academic Publishers

\bibitem[\protect\citeauthoryear{Rackham et~al.,}{Rackham
  et~al.}{2017}]{Rackham17}
Rackham B.,  et~al., 2017, The Astrophysical Journal, 834:151, 21 pp

\bibitem[\protect\citeauthoryear{Rey, Nikitin  \& Tyuterev}{Rey
  et~al.}{2017}]{Rey17}
Rey M.,  Nikitin A.~V.,   Tyuterev V.~G.,  2017, The Astrophysical Journal,
  847:105, 19 pp

\bibitem[\protect\citeauthoryear{Roman \& Rauscher}{Roman \&
  Rauscher}{2019}]{Roman19}
Roman M.,  Rauscher E.,  2019, The Astrophysical Journal, 872, 1

\bibitem[\protect\citeauthoryear{Roman, Kempton, Rauscher, Harada, Bean  \&
  Stevenson}{Roman et~al.}{2020}]{Roman20}
Roman M.~T.,  Kempton E. M.~R.,  Rauscher E.,  Harada C.~K.,  Bean J.~L.,
  Stevenson K.~B.,  2020, arXiv.org, p. arXiv:2010.06936

\bibitem[\protect\citeauthoryear{Rothman et~al.,}{Rothman
  et~al.}{2010}]{Rothman10}
Rothman L.~S.,  et~al., 2010, Journal of Quantitative Spectroscopy and
  Radiative Transfer, 111, 2139

\bibitem[\protect\citeauthoryear{{Sherman}}{{Sherman}}{1963}]{Sherman63}
{Sherman} F.,  1963, A Survey of Experimental Results and Methods for the
  Transition Regime of Rarefied Gas Dynamics.
Academic Press, p.~228

\bibitem[\protect\citeauthoryear{Showman, Fortney, Lian, Marley, Freedman,
  Knutson  \& Charbonneau}{Showman et~al.}{2009}]{Showman09}
Showman A.~P.,  Fortney J.~J.,  Lian Y.,  Marley M.~S.,  Freedman R.~S.,
  Knutson H.~A.,   Charbonneau D.,  2009, The Astrophysical Journal, 699, 564

\bibitem[\protect\citeauthoryear{Sing, Vidal-Madjar, des Etangs, Desert,
  Ballester  \& Ehrenreich}{Sing et~al.}{2008}]{Sing08}
Sing D.~K.,  Vidal-Madjar A.,  des Etangs A.~L.,  Desert J.~M.,  Ballester G.,
   Ehrenreich D.,  2008, The Astrophysical Journal, 686, 667

\bibitem[\protect\citeauthoryear{Sing et~al.,}{Sing et~al.}{2016}]{Sing16}
Sing D.~K.,  et~al., 2016, Nature, 529, 59

\bibitem[\protect\citeauthoryear{Steinrueck, Showman, {Lavvas}, Koskinen, Zhang
   \& Tan}{Steinrueck et~al.}{2020}]{Steinrueck20}
Steinrueck M.,  Showman A.~P.,  {Lavvas} P.,  Koskinen T.,  Zhang X.,   Tan X.,
   2020. No. EPSC2020-768

\bibitem[\protect\citeauthoryear{Swain, Bouwman, Akeson, Lawler  \&
  Beichman}{Swain et~al.}{2008}]{Swain08}
Swain M.~R.,  Bouwman J.,  Akeson R.~L.,  Lawler S.,   Beichman C.~A.,  2008,
  The Astrophysical Journal, 674, 482

\bibitem[\protect\citeauthoryear{Swain, Vasisht, Tinetti, Bouwman, Chen, Yung,
  Deming  \& Deroo}{Swain et~al.}{2009}]{Swain09}
Swain M.~R.,  Vasisht G.,  Tinetti G.,  Bouwman J.,  Chen P.,  Yung Y.,  Deming
  D.,   Deroo P.,  2009, The Astrophysical Journal Letters, 690, L114

\bibitem[\protect\citeauthoryear{Thorngren, Gao  \& Fortney}{Thorngren
  et~al.}{2019}]{Thorngren19}
Thorngren D.~P.,  Gao P.,   Fortney J.~J.,  2019, The Astrophysical Journal
  Letters, p.~L6

\bibitem[\protect\citeauthoryear{Vardavas \& Carver}{Vardavas \&
  Carver}{1984}]{Vardavas84}
Vardavas I.~M.,  Carver J.~H.,  1984, Planetary And Space Science, 32, 1307

\bibitem[\protect\citeauthoryear{Vardavas \& {Taylor}}{Vardavas \&
  {Taylor}}{2007}]{Vardavas07}
Vardavas I.~M.,  {Taylor} F.~W.,  2007, Radiation and Climate.
Oxford

\bibitem[\protect\citeauthoryear{Venot, H{\'e}brard, Ag{\'u}ndez, Decin  \&
  Bounaceur}{Venot et~al.}{2015}]{Venot15}
Venot O.,  H{\'e}brard E.,  Ag{\'u}ndez M.,  Decin L.,   Bounaceur R.,  2015,
  Astronomy and Astrophysics, 577, 12 pp

\bibitem[\protect\citeauthoryear{Visscher, Lodders  \& Fegley}{Visscher
  et~al.}{2010}]{Visscher10}
Visscher C.,  Lodders K.,   Fegley B.~J.,  2010, The Astrophysical Journal,
  716, 1060

\bibitem[\protect\citeauthoryear{Wakeford \& Sing}{Wakeford \&
  Sing}{2015}]{Wakeford15}
Wakeford H.~R.,  Sing D.~K.,  2015, Astronomy and Astrophysics, 573, A122

\bibitem[\protect\citeauthoryear{{West}, {Baines}, {Friedson}, {Banfield},
  {Ragent}  \& {Taylor}}{{West} et~al.}{2007}]{West07}
{West} R.~A.,  {Baines} K.~H.,  {Friedson} A.~J.,  {Banfield} D.,  {Ragent} B.,
    {Taylor} F.~W.,  2007, Jupiter, Jovian Clouds and Haze.
Cambridge, p.~79

\bibitem[\protect\citeauthoryear{{West}, {Baines}, {Karkoschka}  \&
  {S{\'a}nchez-Lavega}}{{West} et~al.}{2009}]{West09}
{West} R.~A.,  {Baines} K.~H.,  {Karkoschka} E.,   {S{\'a}nchez-Lavega} A.,
  2009, {Clouds and Aerosols in Saturn's Atmosphere},
  \mn@doi{10.1007/978-1-4020-9217-6_7.
}

\bibitem[\protect\citeauthoryear{{West}, {Lavvas}, {Anderson}  \&
  {Imanaka}}{{West} et~al.}{2013}]{West13}
{West} R.~A.,  {Lavvas} P.,  {Anderson} C.,   {Imanaka} H.,  2013, Titan.
Cambridge Universtity Press

\bibitem[\protect\citeauthoryear{Woitke, Helling, Hunter, Millard, Turner,
  Worters, Blecic  \& Stock}{Woitke et~al.}{2018}]{Woitke18}
Woitke P.,  Helling C.,  Hunter G.~H.,  Millard J.~D.,  Turner G.~E.,  Worters
  M.,  Blecic J.,   Stock J.~W.,  2018, Astronomy and Astrophysics, 614, A1

\bibitem[\protect\citeauthoryear{Yousefi, Bernath, Hodges  \& Masseron}{Yousefi
  et~al.}{2018}]{Yousefi18}
Yousefi M.,  Bernath P.~F.,  Hodges J.,   Masseron T.,  2018, Journal of
  Quantitative Spectroscopy and Radiative Transfer, 217, 416

\bibitem[\protect\citeauthoryear{Zahnle, Marley, Freedman, Lodders  \&
  Fortney}{Zahnle et~al.}{2009}]{Zahnle09}
Zahnle K.,  Marley M.~S.,  Freedman R.~S.,  Lodders K.,   Fortney J.~J.,  2009,
  The Astrophysical Journal Letters, 701, L20

\bibitem[\protect\citeauthoryear{Zellem et~al.,}{Zellem
  et~al.}{2014}]{Zellem14}
Zellem R.~T.,  et~al., 2014, The Astrophysical Journal, 790:53, 9 pp

\bibitem[\protect\citeauthoryear{Zhang \& Showman}{Zhang \&
  Showman}{2018}]{Zhang18}
Zhang X.,  Showman A.~P.,  2018, The Astrophysical Journal, 866:2, 15 pp

\bibitem[\protect\citeauthoryear{Zhang, Strobel  \& Imanaka}{Zhang
  et~al.}{2017}]{Zhang17}
Zhang X.,  Strobel D.~F.,   Imanaka H.,  2017, Nature, 551, 352

\makeatother
\end{thebibliography}




%
%


\bsp	
\label{lastpage}
\end{document}